\documentclass[fleqn,usenatbib]{mnras}
\usepackage{newtxtext,newtxmath}

\usepackage[T1]{fontenc}

\DeclareRobustCommand{\VAN}[3]{#2}
\let\VANthebibliography\thebibliography
\def\thebibliography{\DeclareRobustCommand{\VAN}[3]{##3}\VANthebibliography}

\usepackage{graphicx}	
\usepackage{amsmath}	

\usepackage{amssymb}	
\usepackage{multirow}
\usepackage{enumitem}
\usepackage{booktabs}
\usepackage{graphicx}
\usepackage{bm}
\usepackage[flushleft]{threeparttable}
\usepackage{mathrsfs}
\usepackage[normalem]{ulem}

\def\lcdm{\ensuremath{\Lambda}CDM}
\def\hc{\ensuremath{H_0}}
\def\tdd{\ensuremath{D_{\Delta t}}}
\def\hst{{\it HST}}

\def\galfit{{\sc Galfit}}

\def\kext{\ensuremath{\kappa_{\rm ext}}}
\def\sersic{S\'ersic}
\def\jcap{Journal of Cosmology and Astroparticle Physics}

\newcommand{\hcuint}{km~s$^{-1}$~Mpc$^{-1}$}
\newcommand{\ucla}{Department of Physics and Astronomy, University of California, Los Angeles, CA, 90095-1547, USA}
\newcommand{\kipac}{Kavli Institute for Particle Astrophysics and Cosmology and Department of Physics, Stanford University, Stanford, CA 94305, USA}
\newcommand{\ucd}{Department of Physics, University of California, Davis, CA 95616, USA}
\newcommand{\tum}{Physik-Department T38, Technische Universit\"at M\"unchen, James-Franck-Str.\ 1, D-85748 Garching, Germany}
\newcommand{\icsuz}{Institute for Computational Science, University of Zurich, 8057 Zurich, Switzerland}
\newcommand{\piuz}{Physics Institute, University of Zurich, 8057 Zurich, Switzerland}
\newcommand{\mpia}{Max Planck Institute for Astrophysics, Karl-Schwarzschild-Strasse 1, D-85740 Garching, Germany}
\newcommand{\epfl}{Institute of Physics, Laboratory of Astrophysics, Ecole Polytechnique F\'ed\'erale de Lausanne (EPFL), Observatoire de Sauverny, 1290 Versoix, Switzerland}
\newcommand{\kipaclm}{Kavli Institute for Particle Astrophysics and Cosmology, Stanford University, 452 Lomita Mall, Stanford, CA 94035, USA}
\newcommand{\ipmu}{Kavli IPMU (WPI), UTIAS, The University of Tokyo, Kashiwa, Chiba 277-8583, Japan}
\newcommand{\stariqa}{STAR Institute, Quartier Agora - All\'ee du six Ao\^ut, 19c B-4000 Li\`ege, Belgium}
\newcommand{\chasc}{International CHASC Astrostatistics Collaboration, Harvard University, Cambridge, MA 02138, USA}
\newcommand{\capsu}{Center for Astrostatistics, The Pennsylvania State University, University Park, PA 16802, USA}
\newcommand{\dspsu}{Department of Statistics, The Pennsylvania State University, University Park, PA 16802, USA}
\newcommand{\daapsu}{Department of Astronomy and Astrophysics, The Pennsylvania State University, University Park, PA 16802, USA}
\newcommand{\icdpsu}{Institute for Computational and Data Sciences, The Pennsylvania State University, University Park, PA 16802, USA}
\newcommand{\tsinghua}{Department of Astronomy, Tsinghua University, Beijing, 100084, China}
\newcommand{\iauc}{Institute of Astronomy, University of Cambridge, Madingley Road, Cambridge CB3 0HA, UK}
\newcommand{\aries}{Aryabhatta Research Institute of observational sciencES (ARIES), Nainital 263002, India}
\newcommand{\cuhp}{Department of Physics and Astronomical Sciences, Central University of Himachal Pradesh, India-176206}
\newcommand{\eub}{Exzellenzcluster Universe, Boltzmannstr. 2, 85748 Garching, Germany}
\newcommand{\lmuus}{Ludwig-Maximilians-Universit{\"a}t, Universit{\"a}ts-Sternwarte, Scheinerstr. 1, 81679 M{\"u}nchen, Germany}
\newcommand{\dpui}{Department of Physics, University of Illinois at Urbana-Champaign, 1110 West Green Street, Urbana, IL 61801, USA}
\newcommand{\spaum}{School of Physics and Astronomy, University of Minnesota, 116 Church Street SE, Minneapolis, MN 55455, US}
\newcommand{\uchicago}{Department  of  Astronomy  \&  Astrophysics,  University  of Chicago, Chicago, IL 606374, USA}
\newcommand{\zentrum}{Zentrum f{\"u}r Astronomie der Universit{\"a}t Heidelberg, Institut f{\"u}r Theoretische Astrophysik, Albert-Ueberle-Str. 2, 69120 Heidelberg}

\title[TDLMC]{Time Delay Lens Modelling Challenge}

\author[X. Ding et al.]{
X. Ding,$^{1,2}$\thanks{dxh@astro.ucla.edu}
T. Treu,$^{1}$
S. Birrer,$^{3}$
G.~C.-F. Chen,$^{4}$
J. Coles,$^{5}$
P. Denzel,$^{6,7}$
M. Frigo,$^{8}$\newauthor
A. Galan,$^{9}$
P. J. Marshall,$^{10}$
M. Millon,$^{9}$
A. More,$^{2}$
A. J. Shajib,$^{1, 11}$
D. Sluse,$^{12}$\newauthor
H. Tak,$^{13,14,15,16,17}$
D. Xu,$^{18}$
M. W. Auger,$^{19}$
V. Bonvin,$^{9}$
H. Chand,$^{20,21}$
F. Courbin,$^{9}$\newauthor
G. Despali,$^{22}$
C. D. Fassnacht,$^{4}$
D. Gilman,$^{1}$
S. Hilbert,$^{23,24}$
S. R. Kumar,$^{20}$
J. Y.-Y. Lin,$^{25}$\newauthor
J. W. Park,$^{10}$
P. Saha,$^{7,6}$
S. Vegetti,$^{8}$
L. Van de Vyvere,$^{12}$
L. L.R. Williams,$^{26}$
\\ \\
$^{1}${\ucla}\\
$^{2}${\ipmu}\\
$^{3}${\kipac}\\
$^{4}${\ucd}\\
$^{5}${\tum}\\
$^{6}${\icsuz}\\
$^{7}${\piuz}\\
$^{8}${\mpia}\\
$^{9}${\epfl}\\
$^{10}${\kipaclm}\\
$^{11}${\uchicago}\\
$^{12}${\stariqa}\\
$^{13}${\chasc}\\
$^{14}${\capsu}\\
$^{15}${\dspsu}\\
$^{16}${\daapsu}\\
$^{17}${\icdpsu}\\
$^{18}${\tsinghua}\\
$^{19}${\iauc}\\
$^{20}${\aries}\\
$^{21}${\cuhp}\\
$^{22}${\zentrum}\\
$^{23}${\eub}\\
$^{24}${\lmuus}\\
$^{25}${\dpui}\\
$^{26}${\spaum}
}

\date{Accepted XXX. Received YYY; in original form ZZZ}

\pubyear{2020}

\begin{document}
\label{firstpage}
\pagerange{\pageref{firstpage}--\pageref{lastpage}}
\maketitle

\begin{abstract}
In recent years, breakthroughs in methods and data have enabled gravitational time delays to emerge as a very powerful tool to measure the Hubble constant \hc. However, published state-of-the-art analyses require of order 1 year of expert investigator time and up to a million hours of computing time per system. Furthermore, as precision improves, it is crucial to identify and mitigate systematic uncertainties. With this time delay lens modelling challenge we aim to assess the level of precision and accuracy of the modelling techniques that are currently fast enough to handle of order 50 lenses, via the blind analysis of simulated datasets
. The results in Rung~1 and Rung~2 show that methods that use only the point source positions tend to have lower precision ($10-20\%$) while remaining accurate. In Rung~2, the methods that exploit the full information of the imaging and kinematic datasets can recover \hc\ within the target accuracy ($|A|<2$\%) and precision ($<6$\% per system), even in the presence of a poorly known point spread function and complex source morphology. A post-unblinding analysis of Rung 3 showed the numerical precision of the ray-traced cosmological simulations to be insufficient to test lens modelling methodology at the percent level, making the results difficult to interpret. A new challenge with improved simulations is needed to make further progress in the investigation of systematic uncertainties. For completeness, we present the Rung~3 results in an appendix and use them to discuss various approaches to mitigating against similar subtle data generation effects in future blind challenges.
\end{abstract}



\section{Introduction} \label{sec_intro}
The flat $\Lambda$ cold dark matter (\lcdm) cosmological model has been remarkably successful in explaining the geometry and dynamics of our Universe. It has been able to predict/match the results of a wide range of experiments covering a wide range of physical scales~\citep{Pla++13, Planck2015XIII, Rie++16, Betoule2014, Eisenstein2005, Alam2017}, and the expansion of our Universe~\citep{Riess1998, Perlmutter1999}. 

One of the key parameters of the model is the Hubble constant (\hc) that determines the age and physical scale of the Universe. Measuring \hc\ at high precision and accuracy has been one of the main goals of observational cosmology for almost a century \citep{Freedman2001}. In recent years, as the precision of the measurements has improved to a few percent level, a strong tension has emerged between early and late universe probes. As far as early-universe probes are concerned, analysis of {\it Planck} data yields $H_0 = 67.4 \pm 0.5$ \hcuint~\citep{Planck2018}, assuming a \lcdm\ model. In the local universe, the Equation of State of dark energy (SH0ES) team using the traditional ``distance ladder'' method based on Cepheid calibration of type Ia supernovae by  finds $H_0 = 74.03 \pm 1.42$ \hcuint~\citep{Riess2019}, and $H_0 = 72.4 \pm 2.0$ \hcuint~ based on the tip of the red giant brand \citep{Yuan2019}. The Carnegie-Chicago Hubble Program calibrated the tip of the red giant branch and applied to type Ia supernovae, finding a midway Hubble tension as \hc $= 69.8 \pm 0.8~(\pm1.1\%~{\rm stat}) \pm 1.7~(\pm2.4\%~{\rm sys})$~\citep{Freedman2019}. The tension between late and early universe probes ranges between 4-6~$\sigma$ \citep[see summary by][]{verdetreuriess2019}. If this $\sim8\%$ difference is real and not due to unknown systematics in multiple measurements, it demonstrates that \lcdm\ is not a good description of the universe, and additional ingredients such as new particles or early dark energy might be needed \cite[e.g.,][]{knox2020,Aarendse2020}. Given the potential implications of this tension,  it is crucial to have several independent methods to measure \hc\ each with sufficient precision to resolve the tension (e.g.,~1.6\% to resolve the 8\% \hc\ tension at 5$-\sigma$).

Time-delay cosmography by strong gravitational lensing provides a one-step measurement of \hc\ together with other cosmological parameters~\citep{Refsdal1966, Treu2016}. The strongly lensed source produces multiple images, corresponding to multiple paths followed by the photons through the universe. According to Fermat's principle, the lensed images arrive at the observer at different times, corresponding to the extrema of the arrival time surface. The time delays between the images depend on the absolute value of cosmological distances, chiefly through the so-called ``time-delay distance", \tdd, and can thus be used to infer \hc\ like any other distance indicator~\citep{Schechter1997, Treu2002a, Suy++10}. Importantly, time delay cosmography is independent of all other probes of \hc.

At the time of writing, the \hc\ Lenses in COSMOGRAIL's Wellspring (H0LiCOW) and SHARP collaborations have finished the analysis of six strong lensed quasars and obtain a joint inference for Hubble constant as $H_0 = 73.3\substack{+1.7\\-1.8}$ \hcuint ~\citep{Wong2019}. In addition, as part of the STRIDES collaboration, \citet{Shajib2019_0408} analyzed one particularly information-rich strong lens system DES J0408-5354 alone and constrained the \hc\ at 3.9\% level, in excellent agreement with the \citet{Wong2019} result. (In the rest of the paper, we refer to H0LiCOW/SHARP/COSMOGRAIL/STRIDES collectively as TDCOSMO~\citep{Millon2019}). Measurements of \hc\ using time delay lenses also have been investigated by other collaborations~\citep{Paraficz2010, Ghosh2020}.

Based on the current results, it is predicted that a 1\% precision in the \hc\ can be achieved via the time-delay cosmography alone using a sample of 40 lensed systems~\citep{Shajib2017}.
However, two issues need to be addressed before a 1\% measurement of \hc\ can be achieved with time delay cosmography. First, the analysis and computational costs need to be reduced in order to make the larger samples tractable. Second, all sources of potential systematic uncertainties must be investigated in order to identify and mitigate any outstanding one. 

The first issue is well illustrated by the current state-of-the-art. At present, the analysis of each system requires approximately one year of effort full time by an expert investigator. Furthermore, the analysis by \citet{Shajib2019_0408}  required approximately 1 million hours of CPU time. Analyzing 40 lenses would thus be prohibitive with current techniques, especially in terms of investigator time.  Efforts are underway to automate these modelling efforts so that they can be scaled to a large number of lenses reducing the investigator time per lens \citep{shajib2019is}, but much work remains to be done to get to high precision, low cost modelling (Schmidt et al. 2020, in prep).

Regarding the second issue, a number of efforts are under way to identify systematic uncertainties \citep[e.g.][]{Millon2019}. All parts of the analysis need to be checked with high-quality data and independent analysis, as well as with simulated datasets. 

One effective strategy to test for unknown systematic errors is to use blind analysis. In the implementation followed by the TDCOSMO collaboration, the inferred values of \tdd\ and \hc\ are kept blind until all coauthors agree to freeze the analysis during a collaboration telecon. The inference is then unblinded and published without modification. One of the goals of the blind analysis is to avoid conscious and unconscious confirmation bias. 

Another powerful strategy is to study systematic errors using realistic simulations, possibly analyzed blindly. Blind analysis of simulated datasets was the strategy of the ``Time Delay Challenge'' (TDC). In the TDC, a so-called ``Evil'' team first simulated a large number of realistic `mock' time delay light curves, including anticipated physical and experimental effects. Then, the ``Evil'' team published the mock data and invited the community to extract time delay signals {\it blindly} using their own method. \citet{TDC2} showed that time delays can be measured from realistic datasets with precision within 3\% and accuracy within 1\%. 

The success of TDC encouraged the community to take on the next step by verifying the precision and accuracy of lens models with a time delay lens modelling challenge (TDLMC), initiated on 2018 January 8th by posting a paper on arXiv~\citep[TDLMC1,][]{ding2018}. The challenge ``Evil'' team simulated 48 systems of mock strongly lensed quasars data and provided access to the data through a weblink to the participating teams (``Good'' teams) to model, blindly{\footnote{For an early implementation of a blind challenge see paper by~\citet{Williams2000}.}}:
\begin{itemize}[noitemsep]
\item \url{https://tdlmc.github.io}
\end{itemize}
The ``Evil'' team produced realistic simulated time-delay lens data including i) \hst-like lensed AGN images, ii) lens time delays, iii) line-of-sight velocity dispersions, and iv) external convergence. After the ``Good'' team submitted their inferred \hc, the performance of the adopted method could be estimated by comparing them with the true values in the simulation.

The number of simulated lensed quasars was chosen to have sufficient statistics to assess the performance at the percent level (7\% expected per system, gives approximately a $\sim1$\% precision on the mean). We stress that this is already a huge sample for current modelling methods, and thus the challenge is exclusively testing ``fast methods''. 
The computational cost of lens modelling is a major hurdle that will need to be overcome in the future; thus TDLMC uses large simulated samples aiming at testing the speed and performance of these ``fast methods''. 

We also note that TDLMC is limited to the study of the lens model accuracy. Other sources of uncertainty are not considered. Therefore ancillary data, including time delay, line-of-sight velocity dispersion, and information of external convergence are provided  unbiased and with true uncertainties.

This paper provides the details of the challenge design that were hidden in 
the challenge opening paper
 \citep[][hereafter: TDLMC1]{ding2018} and presents an overview of the submission results. We encourage the individual ``Good'' teams to submit more detailed papers on their methods and results. The paper is structured as follows. In Section~\ref{sec_description}, we describe the details of the challenge, including hitherto hidden adopted when simulating the sample. Sections~\ref{sec_response} includes the response from the participating teams to this challenge and a brief summary of the method(s)  adopted. The analysis of the submissions for Rung~1 and Rung 2 is presented in Section~\ref{sec_analysis}.
For Rung~3, we discovered post-unblinding that the numerical precision of the ray-traced simulations was insufficient to test lens model methodology at the percent level, making the results from this rung difficult to interpret. Therefore we dedicate a full Section~\ref{sec:rung3} to the subtleties of Rung 3 that will need to be addressed in a future challenge that wishes to adopt numerical simulations of galaxies as a starting point. The results of Rung~3 are given in Appendix~\ref{app:rung3} for completeness, even though the results should be interpreted with caution. We draw some of the implications of the results and discuss our findings in Section~\ref{sec_implications}. Section~\ref{sec_summary} presents a brief summary of the paper. 

\vspace{0.5cm}
\section{Details of the TDLMC challenge design} \label{sec_description}
There are three challenge ladders in TDLMC, called Rung~1, Rung~2 and Rung~3. In addition, an entry-level Rung~0 is also designed for training propose. To ensure that the ``Good'' teams do not infer any information for the previous rung, we reset the \hc\ at each rung. We adopt two independent codes, namely {\sc Lenstronomy}\footnote{\url{https://github.com/sibirrer/lenstronomy}}~\citep{lenstronomy} and {\sc Pylens}\footnote{\url{https://github.com/tcollett/pylens}}~\citep{Auger2011}, to simulate \hst-like lensed AGN images (equally split). This strategy helps us to mitigate the ``home advantage'', if any, in the sense that when ``Good'' team happens to adopt the same code as the one used to generate the simulated images. The use of two independent codes also allows us to estimate numerical uncertainties related to the implementation of the algorithms, if present.

\subsection{Challenge structure} \label{subsec_structure}
The TDLMC begins with Rung 0, consisting of two lens systems -- one {\it double} and one {\it quad}. This training rung aims to ensure that ``Good'' team members understand the format of the data, and avoids any trivial coding errors or mistakes which potentially affect the results of the entire challenge.

Considering that the lens modelling process is usually time consuming, we generated in total of 48 lensing systems, spread over three blind rungs (i.e., Rung~1,2,3. There are 16 systems in each rung). The sample size is small enough to ensure it is tractable by the ``Good'' teams and large enough to explore different conditions with sufficient statistics and uncover potential biases at the percent level. We increase the level of complexity from Rung~1 to Rung~3.

We reveal the details of the simulations for each blind rung in the rest of this section, including the ones which were only known to the ``Evil'' team before unblinding.

\subsection{Details of each Rung} \label{subsec_rungs_details}
For training purpose, Rung~0 was designed to be as simple as possible. Therefore, simple parametrized forms were adopted to describe the surface brightness of the deflector and the source galaxy (i.e., \sersic), and the mass profile (elliptical power-law) of the deflector. The true point spread function (PSF) is given, and external convergence was not considered. The Rung~0 data is released with all the input parameters so that the ``Good'' teams can validate their analysis.

\begin{figure*}
\centering
\includegraphics[width=1\linewidth]{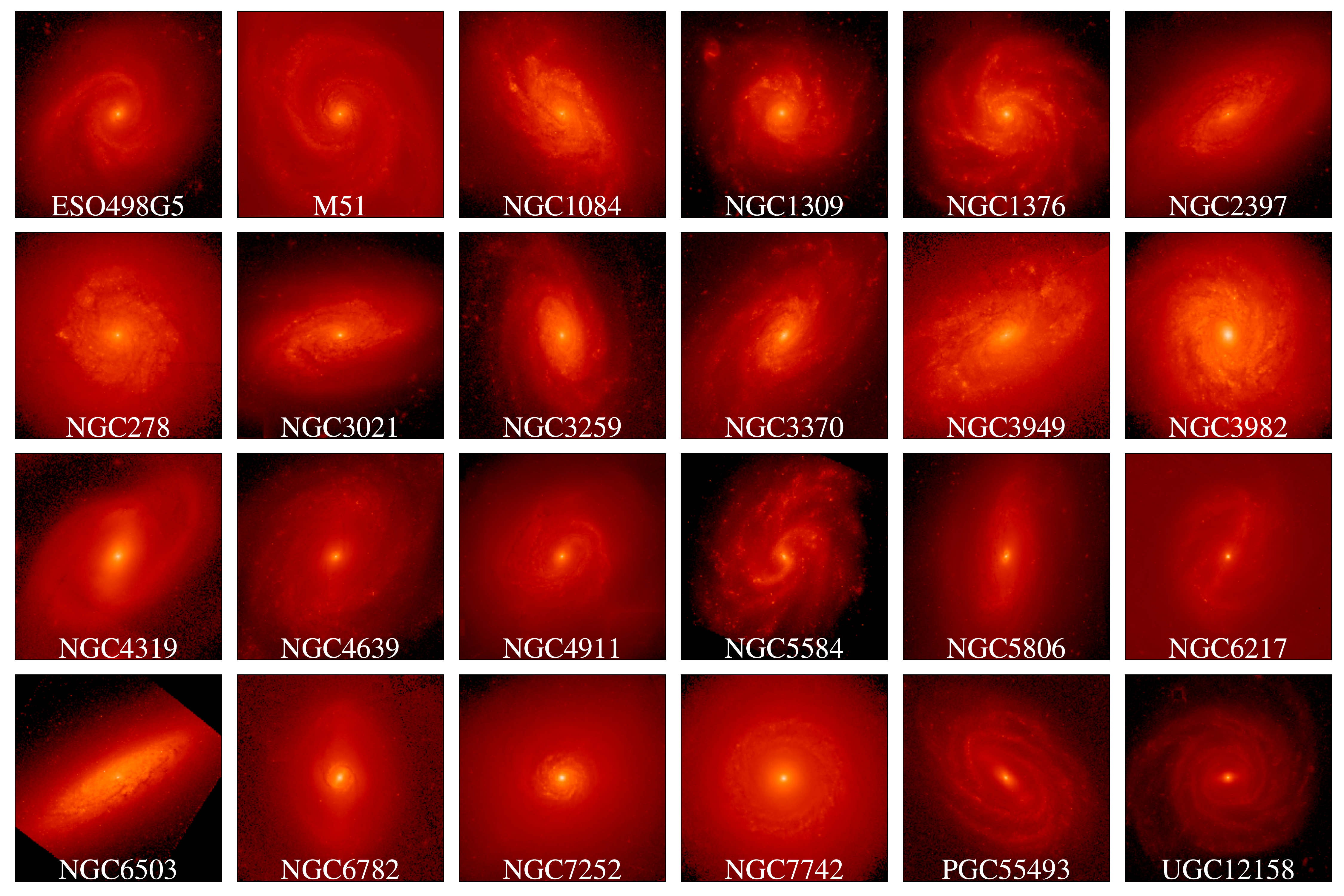}
\caption{
\hst\ images of the realistic galaxies that were used in the TDLMC simulations as lensed AGN host galaxies.
}
\label{fig:source_galaxy}
\end{figure*}

In Rung~1, the increase in complexity with respect to Rung~0 is that the surface brightness of the AGN host galaxy is realistically complex, rather than described by a simply parameterized model like \sersic. For the purpose of making realistic source galaxies, we started from high-resolution images of nearby galaxies obtained by \hst. The digital images are downloaded from the Hubble Legacy Archive\footnote{\url{https://hla.stsci.edu/hlaview.html}}. In order to get a clean galaxy image, we first removed isolated stars and background foreground objects in the field. All the processed galaxy images are shown in Figure~\ref{fig:source_galaxy}.
Then, we obtained the global properties of these galaxies, by using \galfit\ to fit them as the \sersic\ profiles so as to obtain their effective radius ($R_{\rm eff}$) in arcsec and total flux. This information is then used to rescale the galaxy size and magnitude in the source plane, as described in Section~\ref{subsec_sb}. A random external convergence value is also added in Rung~1 (see Section~\ref{subsec_external}).

Rung~2 increases the complexity of Rung~1 by providing the ``Good'' teams with only a guess of the PSF, instead of the actual PSF used to generate the simulations. This added complexity is meant to represent a typical situation where the observer uses a nearby star or model as an initial guess to the actual PSF and then improves on it using the quasar images themselves. In order to implement this step in a realistic manner, the ``Evil'' team took one actual star observed by \hst\ WFC3/F160W and constructed a high-resolution image by interpolation. This PSF image is used to carry out the simulation process described in Section~\ref{subsec_simulations}. However, the PSF information based on a different star was given to the ``Good'' teams.

Rung~3 was the most ambitious as we aimed to increase the complexity of the deflector mass distribution, in addition to retain the complexities of Rung~2. Assessing the effects of the complexity of the deflector mass distribution is crucial to evaluate the performance of modelling methods. For example, the mass sheet degeneracy~\citep[MSD,][]{Falco1985} can be broken by adopting a power-law model to a non-power-law lens mass distribution~\citep{Schneider2013, Schneider2014}. The assumption of any specific mass profile can potentially result in the systematic bias  to the measured Hubble constant, the magnitude of which depends on the difference between the model and the true unknown profile. This effect has been illustrated with cosmological hydrodynamic simulations~\citep{Xu2016, Tagore2018}, suggesting a potential bias could be introduced due to the MSD. In an attempt to model this, the deflector galaxies in Rung~3 are based on cosmological numerical simulations. However, this is also the most conceptually difficult step because we do not have access to the ``true'' mass distribution in real galaxies. For Rung~3, the ``Evil'' team examined two options to produce a  realistic deflector mass. The first option, following \citet{Gilman2017}, is to use the surface brightness distribution of real galaxies, convert it into stellar mass, and add some dark matter components with some prescription. There are challenges to this approach; for example it is not clear how to obtain self-consistent stellar kinematics.  Thus, we discarded this option and decided to~\citep[following e.g.][]{Xu2016} take the results of hydrodynamical simulations as our ``realistic'' mass distribution (specifically Illustris~\citep{Vogelsberger2013, Vogelsberger2014a} and the `zoom' simulations in~\citet{Frigo2019} were adopted). This method has clear advantages but also limitations. For example, the results are only as good in terms of interpreting the real universe as the simulations are accurate, and it is well known that to simulate massive elliptical galaxies accurately is a challenge~\citep[e.g.,][]{Naab2017}. Furthermore, the resolution of the state-of-the-art simulations is finite, and the effects of finite resolution are important at the scale of strong lensing~\citep{Mukherjee2019, Enzi2019}. We did not anticipate additional numerical issues which were discovered post-unblinding and will be discussed later in the paper. 
 
\begin{figure*}
\centering
\begin{tabular}{cc}
\includegraphics[width=0.5\linewidth]{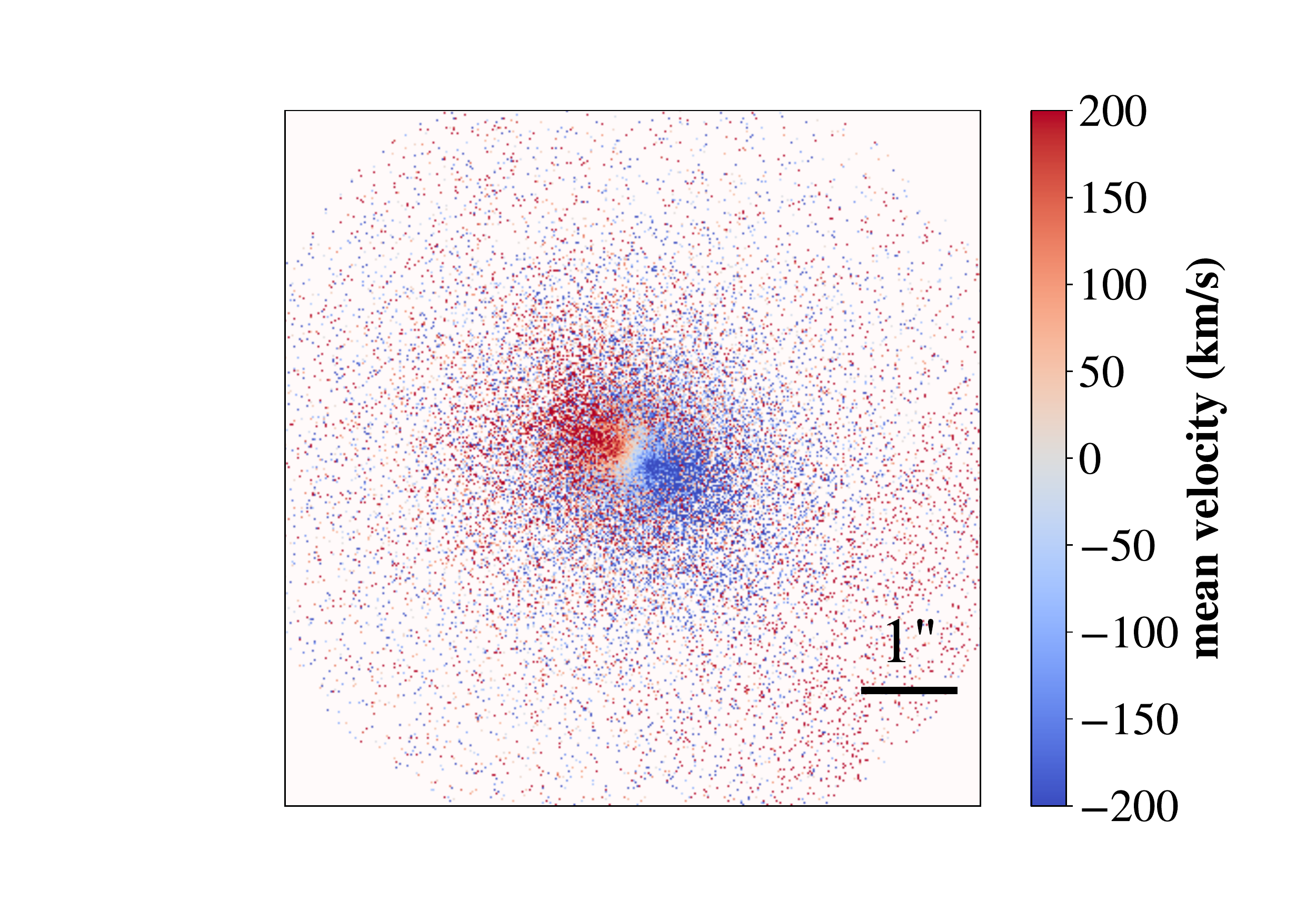}&
\includegraphics[width=0.5\linewidth]{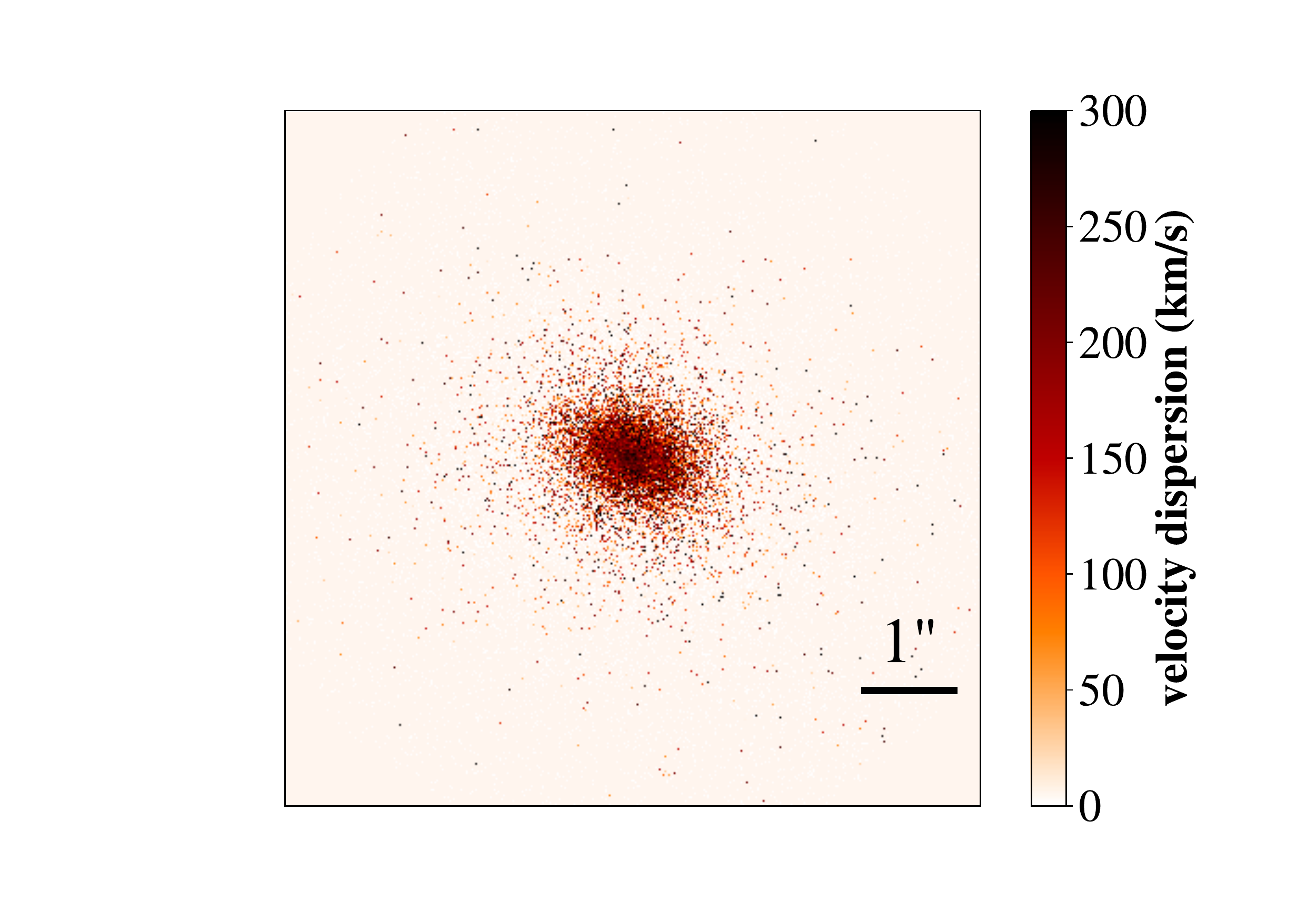}\\
\end{tabular}
\caption{Example of 2D kinematic map for Rung~3. (left): LOS mean velocity ($v_{\rm ave}$) map. (right): velocity dispersion  ($\sigma_{\rm ave}$) map. The image resolution is 16 times higher than \hst/WFC3, i.e., $0\farcs{13}/16 =0\farcs{008125}$. }
\label{fig:vd_map}
\end{figure*}

After setting up the deflector redshift in Section~\ref{subsec_set_redshift}, the Rung~3 deflector providers produced deflector maps at the corresponding redshift. These maps are at very high resolution, which is superior to \hst\ by a factor of 16 (i.e., $0\farcs13/16 = 0\farcs008125$ per pixel). The following information was provided by simulators to generate Rung 3 lensed images including:
\begin{itemize}
\item {\it mass distribution}: The lensing maps include potential map ($f$), the deflection angles maps (including $f'_x$ and $f'_y$, i.e., first-order derivation of $f$) and the hessian map ($f''_{xx}$, $f''_{yy}$ and $f''_{xy}$, i.e., second-order derivation of $f$).
\item {\it surface brightness}: The ``observed'' R-band surface brightness is used to illustrate the light of the deflector in the simulation in Section~\ref{subsec_simulations}. This map is also used to calculate the light-weighted line-of-sight stellar velocity dispersion in Section~\ref{Aperture_vd}. 
\item {\it kinematics}:  The kinematic maps include the line-of-sight averaged velocity map ($V_{\rm ave}$), which accounts for the deflector rotation (Figure~\ref{fig:vd_map}, left panel), and the averaged velocity-dispersion map ($\sigma_{\rm ave}$, see Figure~\ref{fig:vd_map}, right panel).
\end{itemize}

\subsection{Specific ingredients of the simulations} \label{subsec_parameters}
In TDLMC, the ``Evil'' team intends to provide the mock data as realistic as possible. An overview of models/configuration that used to simulate the mock data have been introduced in the challenge designing paper TDLMC1. However, for obvious reasons, some details had to be kept blind and are presented here for the first time.

\subsubsection{Redshift of deflector and source}\label{subsec_set_redshift}
The redshifts of the mock lenses are assumed to be distributed as for typical lenses. In Rung~1 and Rung~2, the ``Evil'' team randomly generated their values from a normal distribution with $z_d \in {\rm N}(0.5 \pm 0.2)$,  $z_s \in {\rm N}(2.0 \pm 0.4)$. In Rung~3, the lensing maps\footnote{Lensing maps include the potential map, the deflection map and the convergence map, i.e., $f$, $f'_x$, $f'_y$, $f''_{xx}$, $f''_{yy}$ and $f''_{xy}$.} are directly provided by the hydrodynamical simulation, fixing the redshift of the source at $z_s = 1.5$ and adopting the same deflector redshift same as provided by the simulation ($z_d\sim 0.5$).

\subsubsection{Detailed setups of the lensing mass}\label{subsec_mass}

The lensing maps are assumed to be composed of a main deflector, plus external shear and convergence. We describe the mass distribution of the deflector in this section. The deflector mass models are meant to describe typical elliptical galaxies.

In Rung~1 and Rung~2, the main deflector is assumed to follow a typical elliptical power-law mass distribution (see also the Section 2.3.1 in TDLMC1), with parameter distributions as listed in Table~\ref{para_config}. In the simulations, we first draw the SIS (i.e., single isothermal sphere) velocity dispersion from the distribution in Table~\ref{para_config}. Then, the corresponding Einstein radius can be calculated as  $R_E = 4\pi v_d^2 \frac{D_{ds}}{D_{s}}$, where $D_{ds}, {D_{s}}$ are the angular diameter distance between the source and the deflector and from the source to us. 

In Rung~3, the deflector mass information is provided by the two simulating teams (X.D., M.F., and S.V.) as described in Section~\ref{subsec_rungs_details}. They also provide the velocity map of the deflector, which is used to calculate the aperture velocity dispersion in Section~\ref{Aperture_vd}, and its surface brightness (see next section).

\subsubsection{Surface brightness calculation}\label{subsec_sb}
The surface brightness in an image is comprised of light both from the deflector and the lensed source.
The main deflector surface brightness in Rung~1 and Rung~2 is described with the widely used \sersic\ profile (as described in Section 2.2.1 in TDLMC1) with parameters distributed as shown in Table~\ref{para_config}. In Rung~3, the (relative) R-band luminosity of stellar particles as deflector light were computed based on their age and metallicity, using the~\citet{Bruzual2003} model.
We only assume the distribution of the deflector's magnitude given in Table~\ref{para_config} to normalize its total flux for the purpose of achieving a realistic signal to noise ratio.

To define the realistic surface brightness distribution of the AGN host galaxy, we adopt a true high-resolution image taken from \hst\ archive for all the blind rungs,. We first rescale the image by projecting it on the source plane, so that it has an apparent \sersic\ effective radius drawn from Table~\ref{para_config}. The magnitude of the source host galaxy is then rescaled from the observed according to the redshift of the source. 

In order to obtain images similar to those used for cosmographic measurements, we assume that the active nuclei have a comparable flux  to that of their host galaxy, see Table~\ref{para_config}.

We vary the position of the source AGN so as to generate the lensing image in a range of configurations (including {\it cusp}, {\it fold}, {\it cross} and {\it double}).

\begin{table}
\centering
\caption{Parameter distributions.}\label{para_config}
\begin{tabular}{ l l}
\hline
Simulation ingredient & model and parameter values \\
\hline\hline  
\\A): redshift \\\hline 
deflector redshift & $z_d \sim {\rm N} (0.5\pm0.2)$ \\ 
source redshift & $z_s \sim {\rm N} (2.0\pm0.4)$ \\ 
 \hline\hline
\\B): deflector (image plane) \\
\hline
lensing galaxy mass & elliptical power-law \\ 
\hline 
SIS velocity dispersion & $v_d\sim {\rm N} (250\pm25)$ km/s \\ 
Einstein radius$^{\rm a}$ &$R_{\rm Ein} = 4\pi v_d^2 \frac{D_{ds}}{D_{s}}$ \\ 
mass slope & $s \sim {\rm N} (2.0\pm 0.1)$ \\ 
ellipticity & $q \sim {\rm U} (0.7 - 1.0)$ \\ 
elliptical axis angle$^{\rm b}$ & $\phi_m \sim {\rm U} (0 - \pi) $ \\  
\hline 
lensing galaxy SB & \sersic\ profile \\ 
\hline 
total magnitude$^{\rm c}$ & $ mag \sim {\rm U} (17.0 - 19.0) $ magnitude \\
effective radius & $R_{\rm eff} = R_{\rm Ein} * {\rm U} (0.5 - 1.0) $ \\ 
\sersic\ index & $n \sim {\rm U} (2.0 - 4.0) $\\
ellipticity & $q \sim {\rm U} (0.7 - 1.0)$ \\
elliptical axis angle$^{\rm d}$ & $\phi = \phi_m {\rm U} (0.9 - 1.1) $ \\
\hline \hline

\\C): AGN (source plane) \\
\hline
host galaxy SB & realistic galaxy\\
\hline
total magnitude & $ mag \sim {\rm U} (22.5 - 20.0) $ magnitude \\
effective radius$^{\rm e}$ & $R_{\rm eff} \sim {\rm U} (0\farcs{}37, 0\farcs{}45), $ {\scriptsize $1.0<z_s<1.5$ \par} \\
&$R_{\rm eff} \sim {\rm U} (0\farcs{}34, 0\farcs{}42), $  {\scriptsize $1.5<z_s<2.0$  \par} \\
&$R_{\rm eff} \sim {\rm U} (0\farcs{}31, 0\farcs{}35), $  {\scriptsize $2.0<z_s<2.5$  \par} \\
&$R_{\rm eff} \sim {\rm U} (0\farcs{}23, 0\farcs{}33), $  {\scriptsize $2.5<z_s<3.0$  \par} \\
\hline
active nuclear light & scaled point source \\
\hline
source plane total flux & $f_{\rm AGN} = f_{\rm host} * {\rm U} (0.8 - 1.25) $ \\
 \\\hline \hline
external shear \\
\hline
amplitudes & $ \gamma\ \sim {\rm U} (0 - 0.05) $ \\
shear axis angle & $\phi \sim {\rm U} (0 - \pi) $ \\  
\hline
external convergency \\
\hline
external kappa$^{\rm f}$ & $ \kappa_{\rm ext} \sim {\rm N} (0 \pm 0.025) $ \\
\hline
\hline
\end{tabular}
  \begin{tablenotes}
 	 \footnotesize
 	 \item Note: $-$ Table lists the assumptions that were used to distribute the parameters for the TDLMC simulation. In Rung~3, non-parameterized deflectors (i.e., lensing galaxy mass and surface brightness) are adopted. Thus, the B part in the table is not adoptable for this rung. The distribution of ``${\rm N}$'' means normal distribution and the ``$ {\rm U}$'' means uniform distribution. Among all the parameters shown in the table, only the redshifts (with zero observation error) and unbiased estimated of external convergence $ \kappa_{\rm ext} = 0 \pm 0.025 $ were provided to the ``Good'' teams.
	 \item ~~~~a: Using our definition, the Einstein Radius would be in the range [1$\farcs$00, 1$\farcs$20]. 
	 \item ~~~~b: The position angles start from the x-axis anti-clockwise.
	 \item ~~~~c: The flux in cps and the magnitude value are related by the equation: \\ ${\rm mag}= -2.5 * \log10({\rm flux}) + {\rm zp}$, where zp is the filter zeropoint in AB system. For filter WFC3/F160W, zp $= 25.9463$.
	 \item ~~~~d: The effective radius and elliptical axis angle of the lensing light are assumed to be correlated with lensing mass at a certain level.
	 \item ~~~~e: The effective radius of the realistic galaxy is obtained by fitting \sersic\ profiles using Galfit.
	 \item ~~~~f: \kext\ is randomly generated to calculate the time delay data. The parent distribution was provided to the ``Good'' teams, but not the actual value, to mimic real analyses, see the descriptions at Eq.~\eqref{eq:k_ext} for more details.
  \end{tablenotes}
\end{table}

\subsection{External shear and convergence}\label{subsec_external}
All the mass along the line-of-sight (LOS) contributes to the deflection of photons. In current state-of-the art analyses, this problem is made tractable by modelling the main deflector and the most massive nearby perturbers explicitly, while describing the remaining effects to first order as external shear and convergence (\kext).

For simplicity, in this challenge we do not include massive perturbers, so there are just two components, the main deflector (described above) and the LOS external shear and convergence.

In Rung~1 and Rung~2, we add an external shear to the lensing potential with typical strength and random orientation, as shown in Table~\ref{para_config}. External shear is not added in Rung~3 in order to keep the lensing potential self-consistent with the mass. 
More important is the effect of the external convergence (\kext), since it affects the relative Fermat potential and time delay. As mentioned in TDLMC1, we consider the effect of \kext\ by drawing from a Gaussian distribution ${\rm N} (0\pm 0.025)$ for all the three Rungs.

\subsection{Generating \hst-like data}\label{subsec_simulations}
Having defined the ingredients of the simulations, we adopt two independent codes to build the pipeline that produces the mock \hst\ imaging data. We aim to simulate the image quality of typical state-of-the-art datasets, i.e., WFC3/F160W with individual exposures of $1200$ s, and typical background. We use astrodrizzle to co-add eight single dithered exposures to obtain the final image with pixel sampling improved from $0\farcs{13}$ to $0\farcs{08}$.

\begin{figure*}
\centering
\includegraphics[trim = 0mm 40mm 0mm 40mm, clip,width=1.0\linewidth]{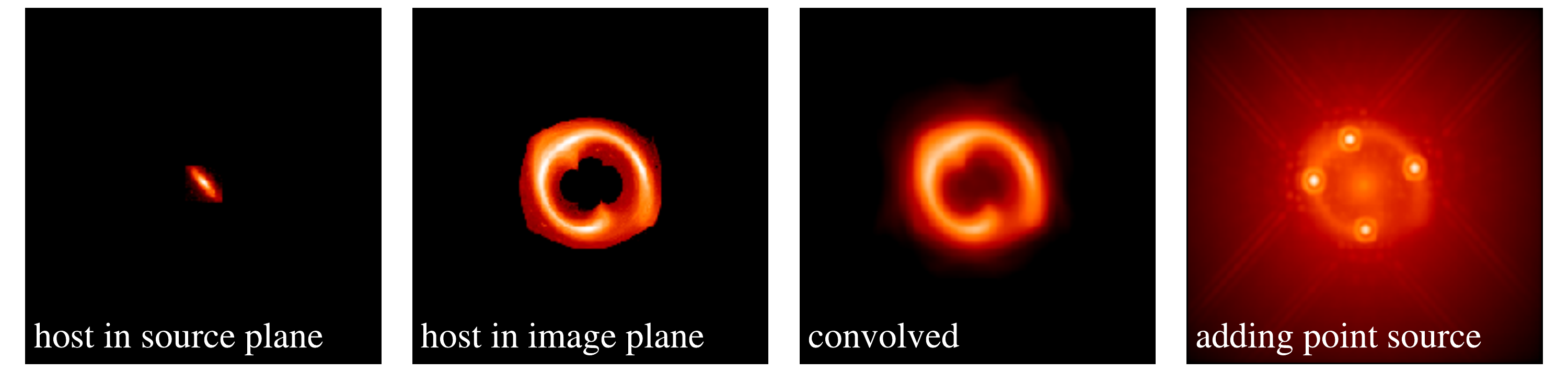}
\includegraphics[trim = 0mm 40mm 0mm 40mm, clip,width=1.0\linewidth]{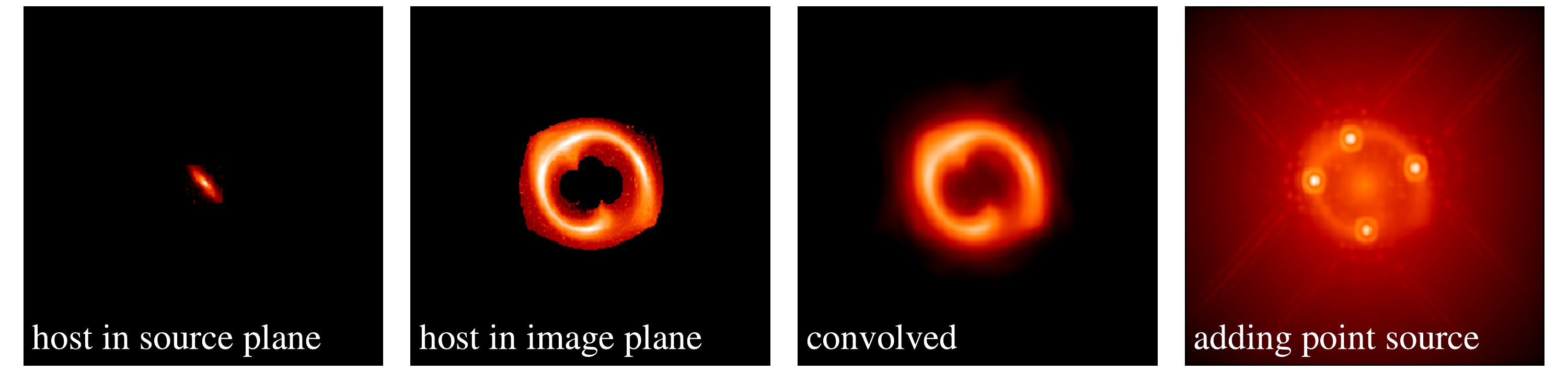}
\caption{
Illustration of the generation of a mock lensed AGN image, using {\sc Lenstronomy} (top) and {\sc Pylens} (bottom). The image is sampled based on \hst-WFC3/F160W at 4 times higher resolution (i.e., $0\farcs{13}/4$). Note that the difference of numerical implementation between the two codes yield little systematic residuals, which is well below the noise level (see Figure~\ref{fig:sim-images}).
}
\label{fig:sim_pipeline}
\end{figure*}

The simulations are similar to those described by~\citet{Ding2017}, which we refer to for more details. A brief description is given here for convenience. The simulation starts from high-resolution images with pixel scale 4 times smaller than the \hst\ resolution (i.e., $0\farcs{13}/4$). We start from actual \hst\ images, as illustrated in Figure~2 of TDLMC1. To numerically define the surface brightness of these actual images, {\sc Pylens} uses interpolation, and with  {\sc Lenstronomy} we chose to use shapelet decomposition \citep{Refregier2003, Birrer2015}. We then rescale the image to the desired size. Then, the distortion by lensing is based on the deflection angles. We convolve the image plane surface brightness with the PSF and add scaled PSF in the position as the point sources to mimic instrumental resolution. In Rung~1 the PSF is generated with {\it TinyTim}~\citep{Krist2011}, while in Rung~2 and Rung~3 PSFs are extracted from the real \hst\ images, and we use interpolation to obtain the PSF image at higher resolution. The pipeline is illustrated in
Figure~\ref{fig:sim_pipeline}. Note that at this step, the images are still sampled at the $0\farcs{13}/4$ resolution.

\begin{figure*}
\centering

\begin{tabular}[b]{c}
\small (a) Rung~1 imaging data \\
\includegraphics[trim = 0mm 20mm 0mm 0mm, clip,width=1.0\linewidth]{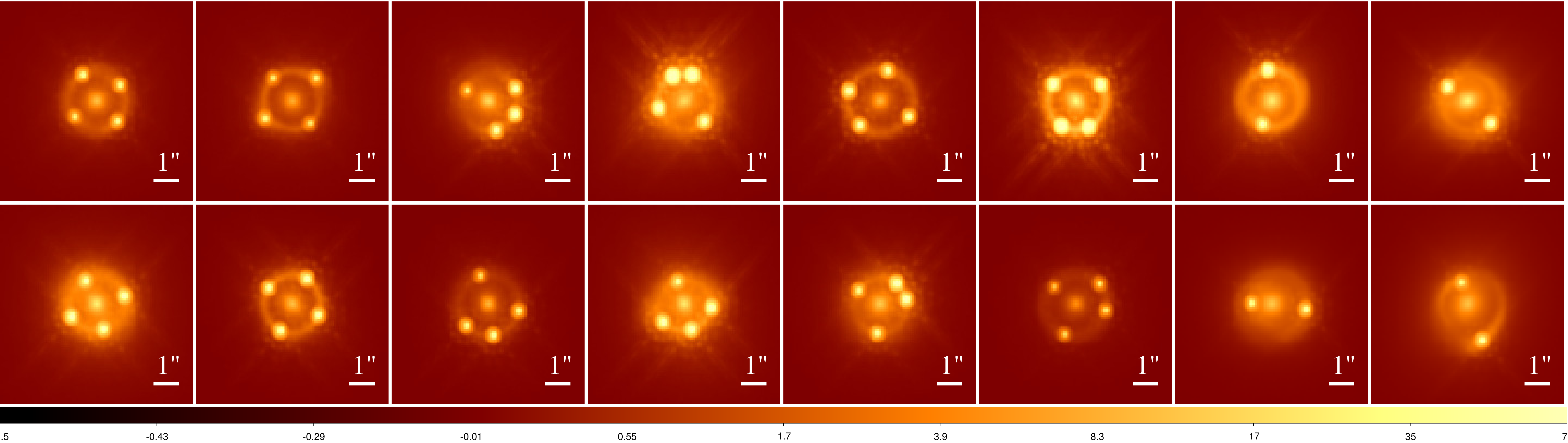}\\
\end{tabular} 

\begin{tabular}[b]{c}
\small (b) Rung~2 imaging data \\
\includegraphics[trim = 0mm 20mm 0mm 0mm, clip,width=1.0\linewidth]{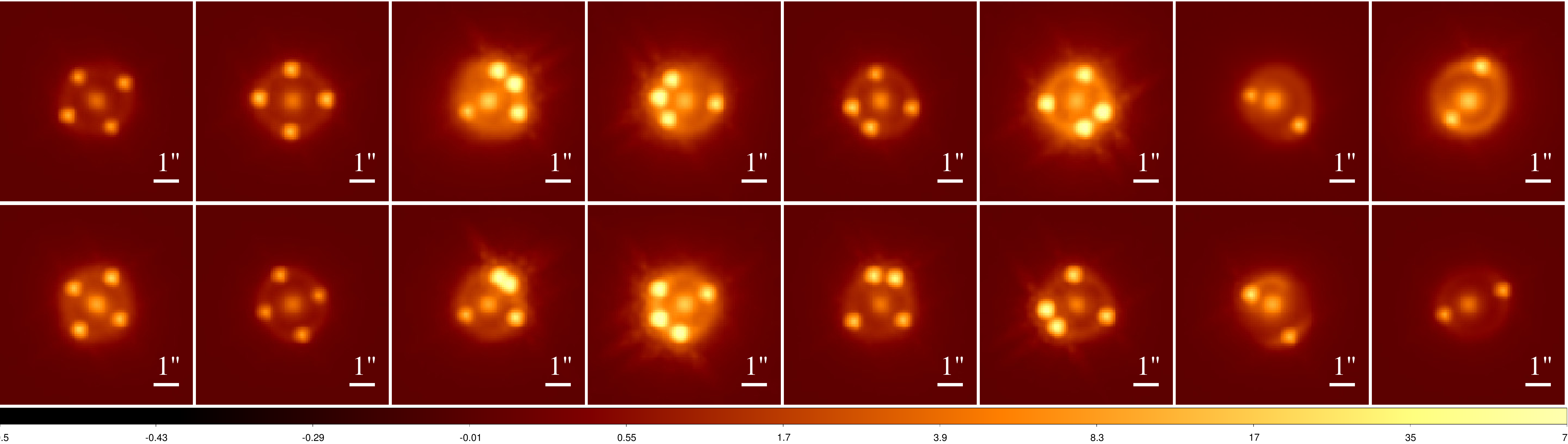}\\
\end{tabular}

\begin{tabular}[b]{c}
\small (c) Rung~3 imaging data \\
\includegraphics[trim = 0mm 20mm 0mm 0mm, clip,width=1.0\linewidth]{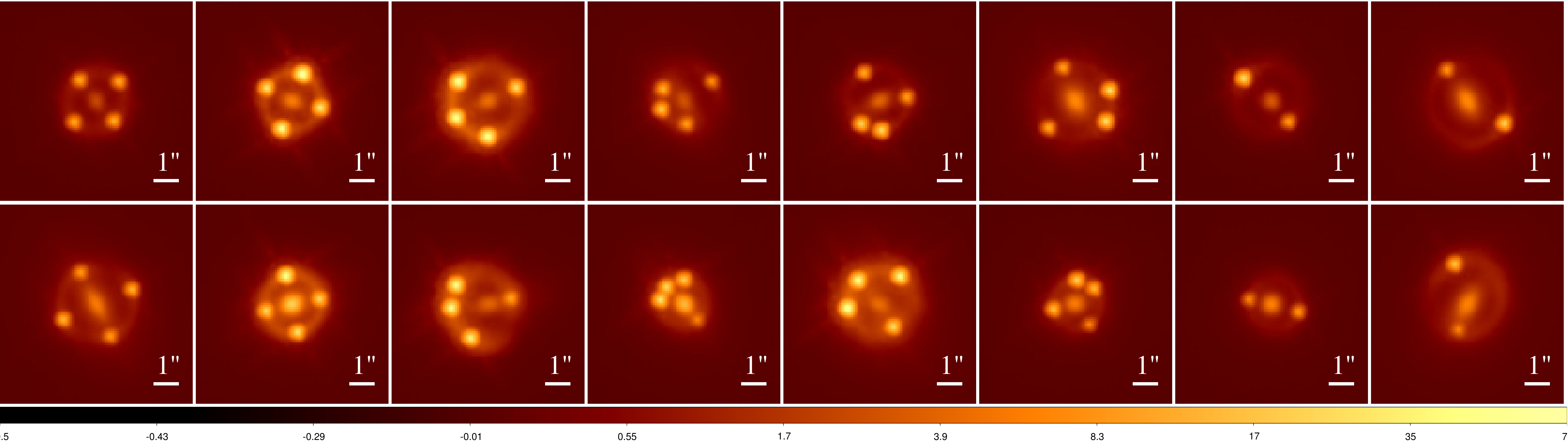}\\
\end{tabular} 

\caption{
Mock data provided for Rung~1, Rung~2 and Rung~3. The configurations from left to right are {\it cross}, {\it cusp}, {\it fold} and {\it double}. The images belonging to the same rung are shown with the same stretch to facilitate visual comparisons.
}
\label{fig:overall_hstdata}
\end{figure*}

In the next step, we rebin the pixels by $4\times4$ to degrade the image at \hst\ resolution, i.e., $0\farcs13$. We select eight different patterns to rebin the image, so as to mimic the dither process. In the next step, we add the noise to the data, see Figure~1 and Figure~2 in~\citet{Ding2017} for details. Finally, we use the drizzling process to co-add eight dithered images to obtain the final drizzled image at $0\farcs08$ sampling. We present the 48 simulated images of the three rungs in Figure~\ref{fig:overall_hstdata}.

In the TDLMC, the eight dithered \hst\ images and the final drizzled images are all provided to the ``Good'' teams including the science images, noise level maps, and a sampled PSF image.

\subsection{Simulated ancillary data} \label{subsec_otherdata}
In addition to the \hst\ imaging data, the ``Evil'' team provides time delay and aperture stellar velocity dispersion, computed as described in this section.

\subsubsection{Time delay}
\label{sec:timedelay}
The {\it true} time delay between the lensed AGN images are calculated using the following equations once the values of the simulated parameters are given by:
\begin{eqnarray}\label{eq:td}
&\Delta t_{ij} = \frac{D_{\Delta t}}{c} \left[
\phi(\theta_i)-
\phi(\theta_j)
\right],
\end{eqnarray}
where ${\theta}_j$ and ${\theta}_j$ are the coordinates of the images $i$ and $j$ in the image plane.
$\phi ({\theta}_i)$ is the Fermat potential at image $i$ and
$D_{\Delta t}$ is so-called time-delay distance, defined as:
\begin{eqnarray}\label{eq:fermat}
& \phi ({\theta}_i)=\frac{({\theta}_i - {\beta})^2}{2} -
\psi({\theta}_i),\\
\label{eq:tdd}
& D_{\Delta t} \equiv (1+z_d) \frac{D_{\rm d} D_{\rm s}}{ D_{\rm ds}},
\end{eqnarray}
where $D_{\rm d}$, $D_{\rm s}$ and $D_{\rm ds}$ are respectively the angular
distances from the observer to the deflector, from the observer to the source, and from the deflector to the source. 

We consider the effects of the \kext\ to the {\rm observed} time delay by:
\begin{equation}\label{eq:k_ext}
\Delta t_{\rm obs} = (1-\kappa_{\rm ext}) \Delta t_{\rm true}.
\end{equation}
Note that the true value of \kext\ is assumed to be zero. It is the measured value of \kext\ that is scattered as ${\rm N} (0\pm 0.025)$. The effect of adding such \kext\ is equivalent to adding a perturbation on the observed time delays as Eq.~\eqref{eq:k_ext}. In principle, the external convergence effect should also shift the Einstein radius, which we did not consider in TDLMC for simplicity. That is, the \kext\ is only taken as a pure scatter effect on the time delay, hence \hc.

Assuming zero bias on the time delay, we add random error as the largest between 1\% and 0.25 days were adopted. We are deliberately keeping the uncertainties on the time delay as small as in the very best cases, in order not to obfuscate lens modelling errors.

\subsubsection{Aperture stellar velocity dispersion}\label{Aperture_vd}

The aperture stellar velocity dispersion is helpful to break the mass sheet degeneracy \citep{Falco1985, Treu2002b}. The integrated line-of-sight velocity dispersion is computed as the second moment of the velocity distribution weighted by surface brightness in a square aperture by $1\farcs0\times1\farcs0$, similar to the standard aperture used for real systems. Seeing conditions are also chosen to mimic the best current ground-based systems, idealized as a Gaussian kernel with a full width at half maximum (FWHM) as $0\farcs$6. 

In Rung~1 and Rung~2 the deflector mass distribution is simply parameterized. Following current practice \citep[e.g.,][]{Shajib2017}, we assume that the mass distribution is related to the velocity dispersion profile through the spherical Jeans equation:
\begin{equation}\label{eq:jeans}
\frac{1}{l(r)}\frac{{\rm d}(l\sigma_{\rm r}^2)}{{\rm d} r} + 2\beta_{\rm ani}(r)\frac{\sigma^2_{\rm r}}{r} = - \frac{GM(\leq r)}{r^2},
\end{equation}
where $l(r)$ is the luminosity density of the deflector galaxy, $\sigma_{\rm r}$ is the radial velocity dispersion and $\beta_{\rm ani}(r)$ is the anisotropy profile and described as:
\begin{equation}\label{eq:jeans}
\beta_{\rm ani}(r) = 1 - \frac{\sigma^2_{\rm t}}{\sigma^2_{\rm r}},
\end{equation}
where $\sigma_{{\rm t}}$ is the tangential velocity dispersion. The observed line-of-sight velocity dispersion is surface-brightness-weighted, and thus can be calculated by solving the equation as~\citet{Mamon2005}
\begin{equation} \label{eq:los-vel-dis}
I(R) \sigma^2_{{\rm los}}(R) = 2 G \int_R^{\infty} {\rm k} \left(\frac{r}{R}, \frac{r_{{\rm ani}}}{R} \right) l(r) M(r) \frac{{\rm d}r}{r}, 
\end{equation}
where $I(R)$ is the deflector surface brightness.

We adopt the Osipkov-Merritt parametrization of anisotropy $\beta_{{\rm ani}}(r) = 1/(1+r_{{\rm ani}}^2/r^2)$~\citep{Osipkov1979,Merritt1985a, Merritt1985b}, with the function ${\rm k}(u, u_{{\rm ani}})$ given by:
\begin{equation}
	\begin{aligned}
	{\rm k} (u, u_{{\rm ani}}) = & \frac{u_{{\rm ani}}^2+1/2}{(u_{{\rm ani}}^2 + 1)^{3/2}}\left( \frac{u^2+u_{\textrm {ani}}^2}{u} \right) \tan^{-1} \sqrt{\frac{u^2-1}{u_{\textrm {ani}}^2+1}} \\
	& \hspace{.1\textwidth} - \frac{1/2}{u_{{\rm  ani}}^2+1} \sqrt{1-1/u^2}.
	\end{aligned} 
	\end{equation}

The anisotropy radius $r_{{\rm ani}}$ is usually considered to be a free parameter with size comparable to the effective radius.  In the simulation, we assume $r_{{\rm ani}} = R_{\rm eff}$ to calculate the velocity dispersion.

\begin{figure}
\centering
\includegraphics[width=1.0\linewidth]{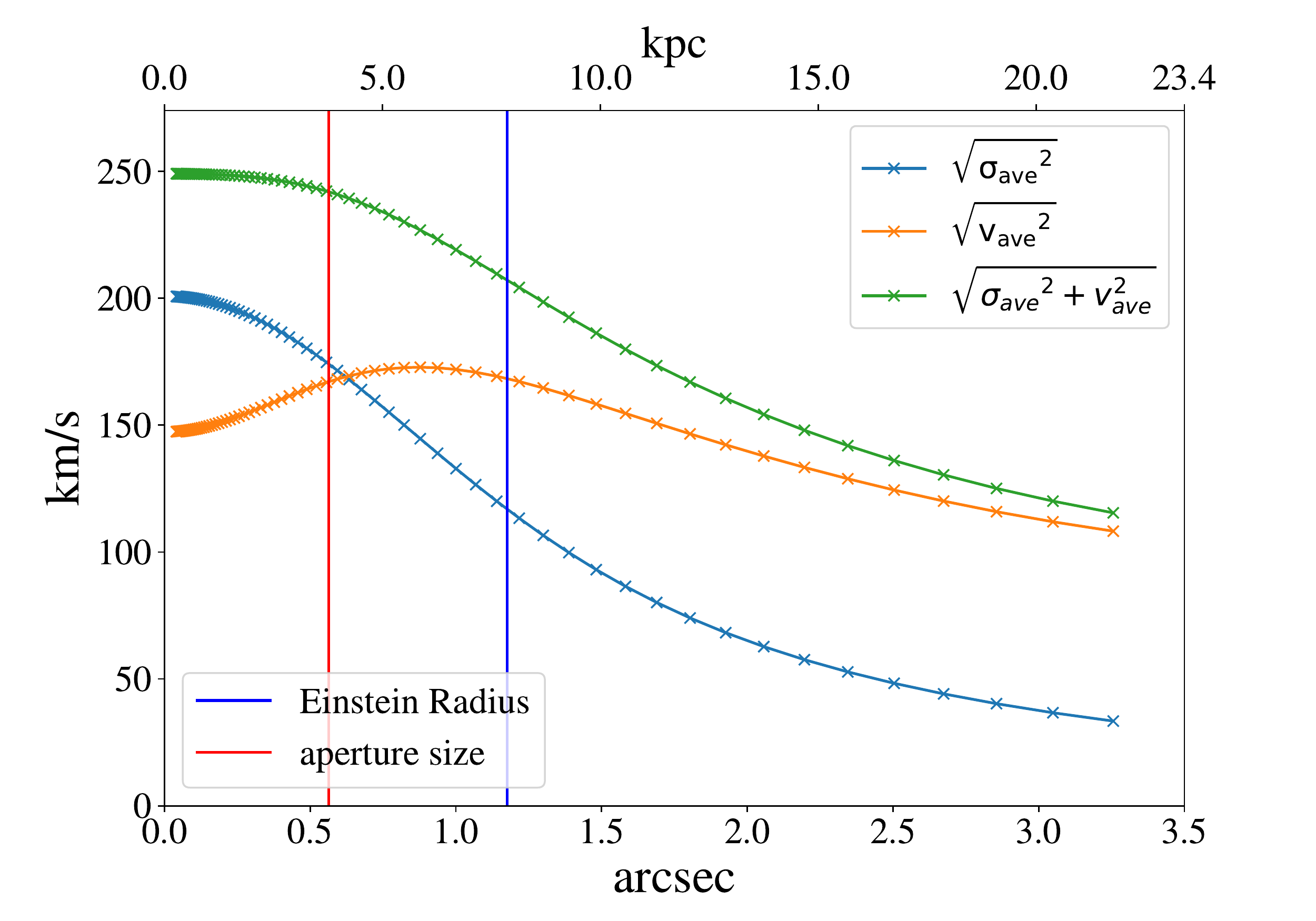}
\caption{
Surface-brightness-weighted line-of-sight stellar velocity dispersion as a function of aperture radius, based on a lens system in Rung~3. The stellar velocity dispersion is composed of the LOS mean velocity (i.e., $v_{\rm ave}$) and LOS velocity dispersion (i.e., $\sigma_{\rm ave}$), added in quadrature. The Einstein radius and the effective aperture size are also shown as blue and red lines, respectively.
}
\label{fig:vd_map_1d}
\end{figure}

In Rung~3, the velocity dispersion is provided by the hydrodynamical simulations via high resolution maps ($16$ times higher than \hst), see Section~\ref{subsec_rungs_details}. The aperture stellar velocity dispersion is thus a combination of the two kinematic maps by: $V_{\rm aper} = \sqrt{V_{\rm ave}^2+\sigma_{\rm ave}^2}$, where the $V_{\rm ave}$ and $\sigma_{\rm ave}$ is the line of sight (LOS) mean velocity and the velocity dispersion as shown in Figure~\ref{fig:vd_map}.
The ``Evil'' team calculate the 2D surface-brightness-weighted line-of-sight dispersion and convolve it using a FWMH $0\farcs$6 Gaussian kernel. Finally, the averaged velocity dispersion in the aperture was computed. Note that in principle the surface brightness weighting should be considered before convolving and aperture selection. However, the velocity map and the surface brightness map are both convolved using the same Gaussian kernel, making the sequence of this processing irrelevant. For illustration, the velocity dispersion as a function of aperture size is shown in Figure~\ref{fig:vd_map_1d}.

A random Gaussian noise with $5\%$ standard deviation is added to the model velocity dispersion to represent high quality measurement errors.

\subsection{Metrics and expected performance} \label{subsec_expectation}

The ``Good'' teams submitted their modelled \hc\ of each lens system in the three rungs, and the ``Evil'' team defined four standard metrics to estimate the performance of the submissions, including {\it efficiency} ($f$), {\it goodness} ($\chi^2$), {\it precision} ($P$) and {\it accuracy} ($A$). They are defined as follows:

\begin{equation}
 \label{eq:efficiency}
f = \frac{N}{N_{\rm total}},
\end{equation} 

\begin{equation}
 \label{eq:goodness}
\chi^2=\frac{1}{N} \sum_i\biggl( \frac{\tilde{H}_{0~i}-H_{0}}{\delta_i} \biggl)^2 ,
\end{equation} 

\begin{equation}
 \label{eq:precision}
P = \frac{1}{N} \sum_i \frac{\delta_i}{H_{0}},
\end{equation} 

\begin{equation}
 \label{eq:accuracy}
A = \frac{1}{N} \sum_i \frac{\tilde{H}_{0~i}-H_{0}}{H_{0}},
\end{equation} 
where $N$ is the number of successfully modelled systems in each submission and $N_{\rm total} = 16$.  ${\delta_i}$ is the uncertainty ($1-\sigma$ level) of \hc\ by each systems in the submission. We identified the following targets for the metrics, based on current state of the analyses:

\begin{eqnarray}
   0.4<\chi^2 <2,\\
   P< 6\% ,\\
   |A| < 2 \%.
\end{eqnarray}

The $\chi^2$ metric target is aimed to ensure that the estimated errors are a reasonable measure of the deviation from the truth. The $P$ metric target is chosen to represent the precision of the best current measurements. The $A$ metric target is set to investigate whether the fast methods can contain biases below the current reported precision by state-of-the-art analysis of samples of a few lenses. We don't set a metric target for $f$, as deciding which systems can be analyzed with sufficient confidence depends on the methodology employed and thus we expect it to vary widely across submissions.

\section{Response to the challenge}\label{sec_response}
The TDLMC challenge mock data were released on 2018 January 8th. The deadlines of the blind submission for the three rungs were: 2018 September 8th for Rung~1; 2019 April 8th for Rung~2; 2019 September 8th for Rung~3. Each rung was unblinded a few days after the submission deadline, to give teams a chance to learn in real time during the challenge. The ``Evil'' team was especially mindful to help the ``Good'' teams detect bugs and glitches that could invalidate the subsequent blind rungs, and prevent the teams from learning about their ability to tackle increased complexity.

Prior to the Rung~3 deadline, the ``Evil'' team received in total 15, 17, and 24 submissions for Rung~1, Rung~2, and Rung~3, respectively, from five different participating teams (``Good'' teams). We describe the method adopted by each team in the rest of this section. 

\subsection{Student-T team} \label{sec:tak}
\begin{center}
\small
H. Tak
\end{center}

This team proposes the following posterior density  of $H_0$ designed to combine information from multiple  lens systems in a simple but statistically principled way:
\begin{equation}\label{tak:eq1}
\pi(H_0 \mid D)\propto L(H_0)h(H_0).
\end{equation}
The notation  $D$ denotes a set of the time delay estimates, Fermat potential difference estimates, and their standard errors for all unique pairs of lenses in 16 systems. Also, $L(H_0)$ represents  the likelihood function  and $h(H_0)$ indicates  a Uniform(20, 120) prior density. This proper uniform prior guarantees posterior propriety of the resulting posterior  \citep{tak2018how}. The team derives the likelihood function from a Gaussian assumption on the Fermat potential difference estimate (Marshall et al., 2020, in preparation):
\begin{equation}\label{tak:eq2}
\phi_{ijk}^{\textrm{est}}\mid \Delta_{ijk}, H_0\sim N\!\left(\phi_{ijk}=\frac{c\Delta_{ijk}}{D_{\Delta}(H_0)},~\sigma^2({\phi_{ijk}^{\textrm{est}}})\right),
\end{equation}
where $\phi_{ijk}^{\textrm{est}}$ denotes the Fermat potential difference estimate of the $i$-th and $j$-th lensed images in the $k$-th lens system and $\sigma({\phi_{ijk}^{\textrm{est}}})$ indicates its standard error ($1\sigma$ uncertainty).  The notation $\Delta_{ijk}$ is the time delay between  the $i$-th and $j$-th lensed images in the $k$-th system ($\Delta_{ijk}=-\Delta_{jik}$). The time delay distance $D_{\Delta}(H_0)$ is treated as a function of only $H_0$ because all other information is completely given in the TDLMC.

On top of this Gaussian assumption on $\phi_{ijk}^{\textrm{est}}$, the team adopts another Gaussian distribution for the  time delay $\Delta_{ijk}$ with its mean equal to ${\Delta}_{ijk}^{\textrm{est}}$ and standard error $\sigma({{\Delta}^{\textrm{est}}_{ijk}})$, i.e.,
\begin{equation}\label{tak:eq3}
\Delta_{ijk}\sim N\!\left({\Delta}^{\textrm{est}}_{ijk},~ \sigma^2({{\Delta}^{\textrm{est}}_{ijk}})\right).
\end{equation}
The team also assumes that $\Delta_{ijk}$ and $H_0$ are independent \emph{a priori} in a sense that $\Delta_{ijk}$ is typically inferred from  light curves of multiply-lensed images without any information about $H_0$ \citep{tak2017bayesian}.

The Gaussian assumptions in Eq.~\eqref{tak:eq2} and \eqref{tak:eq3}  make it simple to integrate out $\Delta_{ijk}$ analytically from their joint distribution, leading to the  Gaussian distribution of ${\phi}^{\textrm{est}}_{ijk}$ given only $H_0$:
\begin{equation}\label{tak:eq4}
{\phi}^{\textrm{est}}_{ijk}\mid H_0\sim N\!\left(\frac{c{\Delta}^{\textrm{est}}_{ijk}}{D_{\Delta}(H_0)},~ \frac{c^2\sigma^2({{\Delta}^{\textrm{est}}_{ijk}})}{D^2_{\Delta}(H_0)}+\sigma^2({{\phi}^{\textrm{est}}_{ijk}})\right).
\end{equation}

The team also assume the conditional independence among Fermat potential difference estimates within and across lensed systems given the Hubble constant $H_0$. Then, the likelihood function of $H_0$ is the product of  Gaussian densities  whose distributions are specified in Eq.~\eqref{tak:eq4}, for every unique pair of  gravitationally lensed images $i$ and $j$ across 16 lensed systems.

Since the posterior density function of $H_0$ in Eq.~\eqref{tak:eq1} is a function of only $H_0$, it is easy to draw an i.i.d.~sample from this posterior via a grid sampling \citep[Chapter~5,][]{gelman2013bayesian}.

On top of the posterior $\pi(H_0 \mid D)$, the team models  \kext\, using  the   relationship,  $H_0^{\textrm{ext}}=(1-\kappa_{\textrm{ext}})H_0$, where $H_0^{\textrm{ext}}$ is the Hubble constant with \kext\ considered  and $H_0$ is the one without \kext\ considered \citep{rusu2017holicow3}. The team puts a $N(0,~ 0.025^2)$ prior  on \kext\ for simplicity, which is assumed to be independent of the data. Finally, the posterior distribution of $H_0^{\textrm{ext}}$ is derived as:
\begin{equation}\label{tak:eq5}
\pi(H_0^{\textrm{ext}}\mid D)=\int \pi(H_0^{\textrm{ext}}\mid D, \kappa_{\textrm{ext}})g(\kappa_{\textrm{ext}})d\kappa_{\textrm{ext}}\ ,
\end{equation}
where $g$ denotes the $N(0,~ 0.025^2)$ density of $\kappa_{\textrm{ext}}$.   The posterior distribution of $H_0^{\textrm{ext}}$ in Eq.~\eqref{tak:eq5} is sampled via a Monte Carlo  integration; (i) draw a random sample of \kext\ from $N(0,~ 0.025^2)$; (ii)  sample $H_0$ from Eq.~\eqref{tak:eq1}; (iii) and lastly set $H_0^{\textrm{ext}}=(1-\kappa_{\textrm{ext}})H_0$. A Jacobian term is not needed for a deterministic transformation within a Bayesian sampling framework \citep{tak2020dta}. The proposed framework does not account for the lens velocity dispersion for each lens system.

The key to the proposed approach is to obtain $D$ to be used as  a condition of the posterior distribution in Eq.~\eqref{tak:eq5} because given $D$, it is simple to draw a random sample of $H_0$. The team notes again that $D$ is composed of time delay estimates, ${\Delta}^{\textrm{est}}_{ijk}$'s, their standard errors, $\sigma({{\Delta}^{\textrm{est}}_{ijk}})$'s, Fermat potential difference estimates, ${\phi}^{\textrm{est}}_{ijk}$'s, and their standard errors, $\sigma({{\phi}^{\textrm{est}}_{ijk}})$'s. The first two components are fully known in the TDLMC, and thus the remaining ingredients for sampling $H_0^{\textrm{ext}}$ from Eq.~\eqref{tak:eq5} are ${\phi}^{\textrm{est}}_{ijk}$'s and $\sigma({{\phi}^{\textrm{est}}_{ijk}})$'s.

For this purpose, the team uses  {\sc Lenstronomy} \citep[version 0.4.3,][]{lenstronomy}. In Rung 1, the team uses the elliptical S\'ersic profile for the source light model and adopts  one, two, or three elliptical S\'ersic profiles for the lens light model. In Rungs 2--3, the team utilizes a superposition of a smooth power-law elliptical mass density profile (SPEMD) with external shear for the lens mass model. An elliptical S\'ersic profile \emph{with} shapelets \citep{Birrer2015} is adopted for the source light model, and an elliptical S\'ersic profile is used for the lens light model. The team fixed $n_\mathrm{max}=10$ as the order of the shapelets basis for the baseline model. Also, the team makes use of the PSF iteration to correct the PSF model  \citep{shajib2019is}. In addition, the team manually boosts the noise level by adopting one of seven different PSF error inflation rates (1\%, 5\%, 10\%, 15\%, 20\%, 25\%,  30\%) to deal with additional errors in the given PSF.  This means that for each unique pairs of lenses, the team fits the  model by {\sc Lenstronomy} seven times each with one of the seven PSF error inflation rates.

For each of the seven fits,  \textsc{Lenstronomy} produces a posterior sample of  ${\phi}_{ijk}$ that is possibly non-Gaussian. Thus, to obtain ${\phi}^{\textrm{est}}_{ijk}$ and $\sigma({{\phi}^{\textrm{est}}_{ijk}})$, the team summarizes the posterior distribution in two ways; posterior mean and standard deviation (Summary~1); posterior median and quantile-based standard error (Summary~2). This is because the posterior mean and standard deviation can be misleading if the posterior distribution of ${\phi}_{ijk}$ is not Gaussian.

Consequently, for each pair of lensed images the team obtains the  seven pairs of $({\phi}^{\textrm{est}(l)}_{ijk}$, $\sigma({{\phi}^{\textrm{est}(l)}_{ijk}}))$ for $l=1, \ldots, 7$, according to each type of summary. Since $D$ requires having only one representative pair of $({\phi}^{\textrm{est}}_{ijk}$, $\sigma({{\phi}^{\textrm{est}}_{ijk}}))$ for each pair of lensed images, the team takes an average of  these seven pairs in three ways. The first one is a Fisher-type weighted average of ${\phi}^{\textrm{est}(l)}_{ijk}$'s weighted by $1/\sigma^2({\phi}^{\textrm{est}(l)}_{ijk})$'s (Average~1). This averaging method puts more weights on the pairs with smaller standard errors.  The second averaging method simply takes an arithmetic mean  over seven estimates and over seven variances (Average~2). This way puts equal weights on all seven pairs regardless of their different standard errors. Finally, the third one uses the same arithmetic mean as Average~2 but sets $\sigma({{\phi}^{\textrm{est}}_{ijk}})$ to a sample variance of the seven estimates, ${\phi}^{\textrm{est}(l)}_{ijk}$'s (Average~3). This one does not use the information about standard errors at all. The team briefly describes the details of each submission in Table~\ref{tak:table}.
\begin{table}
	\centering
	\caption{The details of the submissions of Student-T team. Summaries~1, 2, Averages~1, 2,  3 are defined in Section~\ref{sec:tak}.}
        \label{tak:table}
	\begin{tabular}{ccl} 
		\hline
	Rung	 & Algorithm & Details\\
		\hline
\multirow{7}{*}{$1$} & 1 & Summary~1 and Average~1\\
 & 2 & Summary~1 and Average~2\\
  & 3 & Summary~1 and Average~3\\		
 & 4 & The same as Algorithm~1 except that three  pairs\\
 && are intentionally removed for consistency\\
  & 5 & The same as Algorithm~2 except the three  pairs\\
  & 6 & The same as Algorithm~3 except the three  pairs\\
  \hline
\multirow{5}{*}{$2$}  & 1 & Summary~1 and Average~1\\
    & 2 & Summary~1 and Average~2\\
    & 3 & Summary~2 and Average~1\\
    & 4 & Summary~2 and Average~2\\    
    & 5 & An independent replication of Algorithm~1\\    
   \hline
\multirow{14}{*}{$3$}  & 1 & Summary~1 and Average~1\\
    & 2 & Summary~1 and Average~2\\
    & 3 & Summary~2 and Average~1\\
    & 4 & Summary~2 and Average~2\\    
    & 5 & The same as Algorithm~1 with three times more\\ 
    &&  repetitions (i.e., 21 pairs instead of 7 pairs)\\
    & 6 & The same as Algorithm~2 with 21 pairs\\
    & 7 & The same as Algorithm~3 with 21 pairs\\
    & 8 & The same as Algorithm~4 with 21 pairs\\
    & 9 & The same as  Algorithm~5 but without considering\\
    && \kext\, i.e., sampling from \eqref{tak:eq1} instead of \eqref{tak:eq5}\\
    & 10 & The same as  Algorithm~6 but sampling from \eqref{tak:eq1}\\
    & 11 & The same as  Algorithm~7 but sampling from \eqref{tak:eq1}\\
    & 12 & The same as  Algorithm~8 but sampling from \eqref{tak:eq1}\\
    \hline
	\end{tabular}
\end{table}

Due to the space limitations, the detailed information of the lens modelling settings will be presented in a separate paper (Tak et al., in prep).

\subsection{EPFL team} \label{sec:EPFL}
\begin{center}
\small
M. Millon, A. Galan, F. Courbin, V. Bonvin
\end{center}

\subsubsection{modelling technique}

The EPFL team followed a streamlined version of current modelling practices applied to time delay cosmography. 
The main difference with respect to the analysis described by \citep{Birrer2019_1206,Shajib2019_0408} is that the challenge is known to be free of significant perturbers besides the main deflector and the line of sight. Taking advantage of this information and to reduce computation costs, a smaller number of model choices was considered in the challenge as compared to real systems. In addition, in order to reduce human investigator time, the modelling was standardized as opposed to tailored to the specific of each individual lens. For this purpose, a partly automated modelling pipeline was developed by the team. A more detailed description of the pipeline may be the subject of a future paper. The standardization is a necessary step towards modelling large numbers of systems, but it may result in failures if the one-size-fits all approach is not (yet) sufficiently accurate.

For the modelling part, the team used the publicly available software {\sc Lenstronomy}~\citep{lenstronomy}. This software is well validated and has been previously used for the modelling and cosmography analysis of real time delay strong lens systems \citep{Birrer2016,Birrer2019_1206, Shajib2019_0408}. The entire challenge data set was used as constraints for our models, including the provided drizzled image, noise maps, and PSF ; the measured time delays at lensed AGN positions $\Delta t_\mathrm{measu}$ ; the measured LOS velocity dispersion of stars in the lens galaxy $\sigma_\mathrm{los,\,measu}$ ; the estimate of the external convergence $\kappa_\mathrm{ext}$.

The models are described by linear (surface brightness amplitudes) and non-linear parameters, depending on the type of profiles \citep[see][for details]{Birrer2015}. The team chose to add the time delay distance $D_{\Delta t}$ as a free non-linear parameter.

For a single system, the generic workflow starting from lens modelling up to $H_0$ inference can be divided in the three following steps.

{\bf 1) Parameters optimization and sampling} First linear and non-linear parameters are optimized by alternating Particle Swarm Optimizer (PSO) runs and increments of the complexity of lens models. Parameters are sampled from uniform priors, ensuring that all lenses can be modelled from the same initial set of priors. The time delay distance $D_{\Delta t}$, considered as a free non-linear parameter of the model, is constrained by the measured time delays $\Delta t_{ij,\mathrm{measu}}$ by enforcing the modelled time delays to be compatible with the measured ones. Modelled time delays $\Delta t_{ij,\mathrm{model}}$ are computed as follows:
\begin{align}
    \label{eq:epfl:td_ddt}
    \Delta t_{ij,\mathrm{model}} = (1+z_\mathrm{d})\, \frac{D_{\Delta t}}{c} \Delta \Phi_{ij,\mathrm{model}}\ ,
\end{align}
where $z_\mathrm{d}$ is the lens redshift, $\Phi_\mathrm{model}$ is the model Fermat potential, $c$ is the speed of light, and ``$ij$'' defines the difference of the indicated quantity evaluated at the positions of two lensed AGN $i$ and $j$. This procedure gives best fit estimates of the linear and non-linear parameters, that are then used as a starting point of a MCMC sampling. Both PSO and MCMC routines are implemented in {\sc Lenstronomy}, based on the \textsc{CosmoHammer} package \citep{Akeret2013} and \textsc{emcee} \citep{ForemanMackey:2012p12693}.

{\bf 2) Kinematics and angular diameter distances} For each MCMC sample, the team derived in a post-processing step the LOS velocity dispersion $\sigma_\mathrm{los,\,model}$ from model parameters. The team used the Osipkov-Merritt model to solve the spherical Jeans equation, again following current practices e.g., \citet{Suy++10,Shajib2017}, with routines implemented in \textsc{Lenstronomy}. The team computed angular diameter distances from both kinematics and time delays. The sampled time delay distance gives directly the distance ratio $D_\mathrm{d}D_\mathrm{s}/D_\mathrm{ds}$. The modelled LOS velocity dispersion, along with the model parameters $\boldsymbol{\xi}_\mathrm{model}$, are used to compute the distance ratio $D_\mathrm{s}/D_\mathrm{ds}$ from the following relation \citep{Birrer2016}:
\begin{align}
\label{eq:epfl:sigma_los}
    \sigma^2_\mathrm{los,\,model} = \frac{D_\mathrm{s}}{D_\mathrm{ds}} \, c^2\, J(\boldsymbol{\xi}_\mathrm{model},\, r_\mathrm{ani})\ ,
\end{align}
where $J$ captures all dependencies on model parameters and kinematics anisotropy, moving any dependencies on cosmological parameters in the distance ratio. The external convergence was also sampled as $\kext\backsim\mathcal{N}(0,\,0.025)$, to simulate a correction to the time delay distance by any mass external to the main deflector, through:~$D_{\Delta t,\,\mathrm{eff}} = D_{\Delta t}/(1-\kext)$. From the two distance ratios described above, is is straightforward to extract the angular distance to the deflector, namely $D_\mathrm{d}$.

{\bf 3) Cosmography inference for an individual system}
Following \citet{Birrer2019_1206}, the inference of the Hubble constant is performed in the 2D plane defined by angular distances $D_{\Delta t,\,\mathrm{eff}}$ and $D_\mathrm{d}$. This plane encodes the joint constraints from imaging data, time delays, external convergence and lens kinematics. In order to approximate the full covariance between the two $D_{\Delta t,\,\mathrm{eff}}$ and $D_\mathrm{d}$ posteriors, both distributions are used to evaluate the likelihood when inferring $H_0$. Since $\Omega_\mathrm{m}$ is fixed in this challenge, the only cosmological parameter being sampled is the Hubble constant.

{\bf 4) Joint cosmology inference for an entire rung}
The team computed the final inferred $H_0$ value and associated uncertainty estimates for an entire rung in two steps. First, an outlier rejection scheme was performed, according to the following criteria, that were found to be good markers of poor models:
\begin{itemize}
    \item Each individual $H_0$ median value must be inside the prior bounds defined by the TDLMC, i.e., inside $[50,\,90]$ km\,s$^{-1}$Mpc$^{-1}$;
    \item The sampled time delay distance $D_{\Delta t}$ (free parameter constrained by the lens model and time delays) and the modelled time delay distance $D_{\Delta t,\,\mathrm{model}}$ (obtained through Eq.~\eqref{eq:epfl:td_ddt} inversion) must be consistent with each other at the $\lesssim 1\sigma$ level;
    \item The modelled lens velocity dispersion $\sigma^2_\mathrm{LOS,\,model}$ must be consistent at $\lesssim 2\sigma$ level with the measured value;
    \item Each individual $H_0$ posterior must be consistent with each other at the $\lesssim 2\sigma$ level.
\end{itemize}
When all the above criteria were fulfilled, the team kept the model for the joint inference over the rung, for a given model family. This leads to a set of $D_{\Delta t}$ and $D_\mathrm{d}$ pairs of posteriors. The team then performed two joint inferences using:
\begin{itemize}
    \item Only time-delay information. $H_0$ is sampled according to the ensemble of $D_{\Delta t}$ posteriors only. 
    \item Both time-delay and kinematics information. This follows the approach described in \cite{Birrer2019_1206}, $H_0$ is sampled in the 2D plane over the set of $D_{\Delta t}$ and $D_\mathrm{d}$ posteriors. This last option is the standard procedure used for joint inference of real lenses \cite[e.g, ][]{Wong2019}
\end{itemize}
Note that even in the first case of inference $H_0$ from $D_{\Delta t}$ only, knowledge about kinematics still plays a (smaller) role, because of model selection steps are performed \emph{before} the inference.

The joint $H_0$ posteriors described above are computed under the assumption that the systems do not share systematic errors. If this assumption breaks, then one should marginalize from individual distributions, instead of the joint inference. For this reason, the team also submitted $H_0$ posteriors that are marginalized over the selected models. Additional details specific to each rung are given in the following subsections.

\subsubsection{Rung 1}
\
In Rung~1, lens mass and light profiles are simply-parametrized. Hence the team used power-law elliptical mass distribution (SPEMD) \citep{Barkana1998} with external shear profiles to describe the projected mass distribution, and a single \sersic\ profile for the lens surface brightness. For the source, the team used a \sersic\ profile superimposed to a set of shapelets \citep{Ref03a,Birrer2015}. The team chose $n_\mathrm{max}=8$ as the maximum order of the shapelets basis for their the baseline model. When significant residuals were observed at Einstein ring location, $n_\mathrm{max}$ were slightly increased, typically up to $n_\mathrm{max}=14$. The source galaxy centroid (\sersic+shapelets) was fixed to the position of the quasar, itself modelled as a single point source constrained by enforcing lensed images to trace back to the same position in source plane.

The ``Evil'' team kept secret any details related to kinematics modelling assumptions, including the anisotropy model they used for computing velocity dispersion. As stated above, the EPFL team used Osipkov-Merritt modelling for computing velocity dispersions \citep{Osipkov1979,Merritt1985}. This model assumes a parametrized anisotropy parameter $\beta_\mathrm{ani}=r^2/(r^2+r_\mathrm{ani}^2)$, where $r_\mathrm{ani}$ is the anisotropy radius, which defines the radius at which stellar orbits go from being radial (near the center) to isotropic (equally radial and tangential). Standard practices are to sample the anisotropy space through a uniform prior on the anisotropy radius, see e.g., \citet{Suy++12,Shajib2017}. In Rung 1, the team used a uniform prior $r_\mathrm{ani}\backsim\mathcal{U}(0.5,\,5)\,r_\mathrm{eff}$, where $r_\mathrm{eff}$ is the half-light radius of the lens.

The unblinding of Rung~1 revealed that the team's submitted inference was strongly affected by one (or several) systematic error(s), as quantified by an accuracy of $A = 7.512\%$. The main origin of this bias was found to be a consequence of the high precision of measured time delays, which surpasses those of real time delay lenses so far, combined with small angular separation between lensed images. Indeed image separations are on average $\sim 1''$, and time delays are of the order of dozens of days with precision 0.25 days. Typical lensed systems modelled by the TDCOSMO collaboration have on average image separations of $\sim 2.5''$ with time delays precision up to a couple of days. A particularly high precision is therefore required when modelling the position of each lensed images in the setting of the challenge, which is not the case for all real systems analyzed so far. A lack of precision can propagate to a significant bias on the Hubble constant. The bias they observed in their initial Rung~1 submission allowed them to highlight such a requirement, which have been the topic of a dedicated paper by \citet{birrer2019astrometric}. The authors introduced simple formulae that, given an expected precision on the Hubble constant, can be at first order used to estimate the astrometric requirements that must be fulfilled, from image separations and time delays precision. They refer the reader to that paper for consequences of such requirements and quantitative examples. As discussed in Section~\ref{subsec_lesson0}, the problem was solved by the EFPL team by introducing in \textsc{Lenstronomy} a nuisance parameter to describe the unknown difference between true and measured image positions and marginalizing over it.

For Rung~1, the team submitted a single sample of models, and related joint Hubble value, following the description above.

\subsubsection{Rung 2}

In Rung~2, only a guess of the PSF was provided, in order to test PSF reconstruction algorithms. The team used the iterative PSF reconstruction originally implemented in {\sc Lenstronomy}. For a set of baseline models, the team incorporated this routine during parameter optimization, effectively alternating between PSO and PSF reconstructions. Having noticed that the PSF was degraded the same way for each of the 16 lenses of Rung~2, the team computed a median stacked PSF kernel from their best reconstructed kernels. This reconstructed PSF was then used for all of their subsequent Rung~2 modelling attempts. 

Based on Rung~1 knowledge, the team took into consideration the astrometric requirements described in previous subsection, in order to mitigate a potential bias on the inferred Hubble constant. The team allowed extra degrees of freedom to model any unknown uncertainty on the position of AGN images (a.k.a. point sources), by introducing in the parameter space, two new ``offset'' parameters, $\delta_x$ and $\delta_y$, for each of the 2 or 4 images independently. These offsets actually represent the error between the (modelled) position of point sources on the image, and the (predicted) positions at which the Fermat potential is evaluated for time delays computation. These additional parameters are sampled as non-linear parameters, and constrained by time delays and imaging data. The team regularly checked that those offsets were correctly constrained, with amplitudes expected to be below the image pixel scale.

After careful analysis of post-unblinding or Rung 1, the team realized that most consistent results were obtained when $r_\mathrm{ani}\approx r_\mathrm{eff}$. Consequently, in Rung 2, the team fixed the anisotropy radius $r_\mathrm{ani}$ to be equal to the lens half-light radius for all the remaining submissions.

The remaining volume of the parameter space (mass and light profiles of the lens galaxy, light profiles of source galaxy, and quasar model) was identical to those of the previous rung.

The team submitted 4 model samples and corresponding joint value for Rung~2:
\begin{itemize}
    \item \texttt{DdDdt}: the inferred $H_0$ was obtained through joint inference in the 2D plane $\left\{D_{\Delta t,\,\mathrm{eff}},\, D_\mathrm{d}\right\}$~; 
    \item \texttt{margDdDdt}: same as \texttt{DdDdt}, except that the inferred final value was obtained by marginalization over individual $H_0$ posteriors, as opposed to a joint inference~;
    \item \texttt{Ddtonly}: same as \texttt{DdDdt}, except that $H_0$ values were inferred only from the time delay distance $D_{\Delta t,\,\mathrm{eff}}$~;
    \item \texttt{margDdtonly}: same as \texttt{Ddtonly}, except that the inferred final value was obtained by marginalization over individual $H_0$ posteriors.
\end{itemize}

\subsubsection{Rung 3}

For Rung~3, the team used the exact same PSF reconstruction method as for Rung 2. For lens models, they followed the practices of the TDCOSMO collaboration, in the sense that they chose two families of models: power-law and composite. The former consists of elliptical power-law mass distribution with external shear, whereas the latter distinguishes the baryonic mass and dark matter, in addition to the external shear. For the baryonic matter they used a double Chameleon profile \citep[see][for definition]{Suyu2014chameleon} to fit the lens surface brightness, and convert it to surface mass density through a constant mass-to-light ratio, introduced as a free parameter. They modelled the dark matter component as a single elliptical NFW profile.

In order to improve their efficiency in modelling Rung 3 with two model of families, which require significant amount of work, they also used double Chamelon profiles to describe the lens light in their power-law models. This allowed them to extract best fit lens light parameters from their power-law models, and properly initialise the corresponding composite models, for a given lens. Note that it is different than the usual TDCOSMO procedure, where the surface brightness of the lens galaxy is fitted with double S\'ersic for power-law mass models. They checked that no systematic errors were introduced when using double S\'ersic instead of double Chameleon profiles, which is expected as the latter is designed to be a good approximation of the former.

The rest of the procedure was similar to their submissions for Rung 2 and 3, in terms of selection criterions and joint inference. The selection was performed independently for the two model families described above, meaning that their composite and power-law submissions did not necessarily consist in the same modelled lenses, nor the same number of lenses. For each model family, they submitted two submission pairs, with $H_0$ inferred from: 1) joint $\left\{D_{\Delta t,\,\mathrm{eff}},\, D_\mathrm{d}\right\}$ inference, 2) $D_{\Delta t,\,\mathrm{eff}}$ only. Additionally, they submitted a third pair of submissions with a subset of lenses whose models were coincidentally accepted with both model families, which enabled them to combine their inferences from power-law and composite models. More precisely, for a given lens, they marginalised over the two model families, prior to the final joint inference $H_0$ among the different lenses. To summarize, one ended up with 6 submissions for this rung.

\subsection{Freeform team} \label{sec:freeform}
\begin{center}
\small
P. Denzel, J. Coles, P. Saha, L. L.R. Williams
\end{center}
The lenses were reconstructed with the codes GLASS by \cite{GLASS} and its precursor PixeLens by \cite{PixeLens} which are based on the free-form modelling technique.  
In contrast to other methods, free-form lens reconstructions are not restricted to a parametrized family of models, but rather build a lens as superpositions of a large number of mass components, e.g., mass tiles or pixels, with minimal assumptions about the form of the full lens.
The price to pay for the flexibility is that the free parameters outnumber the constraints and thus regularization needs to be imposed to avoid overfitting the data.

While GLASS and Pixelens are completely separate codes, implemented in different languages, and using different Monte Carlo sampling engines, they both share the same approach to free-form lenses. 
Represented as a discrete grid of pixels, the lens potential takes the following form:
\begin{equation}
\label{eq:free-form-potential}
  \psi(\theta) = \sum \kappa_{n} \nabla^{-2}Q_n({\theta}),
\end{equation}
where $\kappa_n$ is the density of the $n$-th mass tile and $Q_n(\theta)$ is the shape integral over the $n$-th pixel.  
Each tile is a square and its contribution $\kappa_{n}Q_n({\theta})$ to the potential at ${\theta}$ can be worked out analytically \citep{Abdelsalam}.
In both GLASS and PixeLens the tiles cover a circular area centered on the lensing galaxy.
The radius of this area $r_p$, in pixels, determines the resolution of a model.
For instance, $r_p=8$ places one tile at the center and eight tiles extending left and right (17 pixels side to side) with a total of 225 pixels covering the entire circular area.
The tile size in arcseconds can be set explicitly or estimated such that there are several rings of pixels outside the outermost image.
Mass distributions that are assumed to be radially symmetric (doubles and some quads) are constrained to have diametrically opposite pixels of equal value, which reduces the number of pixels by half.
GLASS also allows for the central pixel to be further subdivided into $3\times3$ or $5\times5$ sub-pixels, to capture a steeply rising cusp.
In this paper we denote the use of the subdivision with the parameter $sp=3$ or $sp=5$, respectively. 
A central pixel with no subdivision is equivalent to $sp=1$.
Both codes ensure a small region of ``pixel rings'' outside the outermost image.

Quasar image positions, time delays, and redshifts are the only data input for the models.
Image parities are also given but are determined solely from experience and by generating test models to verify image parity assignment.
As is well-known, images are located at extrema of $\nabla\psi$ and the sign of $\nabla\nabla\psi$ determines the parity.

This input is used to create a system of equations which are linear in the source position ${\beta}$ and mass tiles $\kappa_{n}$. 
The intrinsic and well-known problem of lensing arises from the fact that there are infinitely many solutions to these linear equations.  
Free-form techniques usually sample from that solution space according to a few reasonable priors.
Most notably they require non-negative mass tiles, limited to twice the average of all neighboring tiles, and the local density gradient to point typically 45$^{\circ}$ from the center; additionally, the azimuthally-averaged mass profiles must not increase, which still allows for twisting isodensity contours and significantly varying ellipticities with radius.
These priors ensure some minimum level of physical correctness where the density of the reasonably smooth lensing mass is increasing towards the center.
From the information provided by the ``Evil'' team for each rung, further physical parameters and priors could be included:
\begin{itemize}[leftmargin=*]
  \item Redshifts set the distance scales (assuming a standard cosmology of $\Omega_m=0.27$ and $\Omega_\Lambda=0.73$).
  \item The models allowed for external shear.
  \item Time delays were constrained, for GLASS with uncertainties of $\pm 0.25$~days,
    for PixeLens without.
  \item The range of \hc\ was limited to $50-90$~\hcuint.
\end{itemize}
The velocity dispersion information was not used to constrain the models, but can be derived from the models following \cite{Leier09}.

A free-form lens model consists of an ensemble of models; $\sim$1000 typically provide a good cover of the solution space.
A single model may contain more than one lensing system, in which case they are coupled by the requirement that \hc\ must be the same for all systems.

An ensemble usually includes many different convergence maps some of which are unphysical at times.
Generally this is not a problem, as the ensemble average\footnote{Due to the linear nature of the lens equation, a superposition of solutions also is a solution.} washes out these outliers.
Nevertheless, the ensemble can be filtered according to different criteria in order to optimize the ensemble average. 
In Rung~2 for instance, we applied such a post-processing filter based on a simplified version of the source mapping algorithm described in \cite{Denzel20}.  
Instead of only using quasar image positions, the entire photometric information was used to select the most probable models in the following manner.
A $\chi^{2}$ value was computed for each lens model of the ensembles by fitting a synthetic image using the drizzled image data (including science images, noise level maps and a sampled PSF image, while masking out the lensing galaxies in the center). For each ensemble, 300 models with the best values were retained to estimate $H_{0}$. This ensured that only the models which best fit the entire image data were used to infer $H_{0}$. Despite slight improvements on $H_{0}$ the filter was abandoned again for Rung~3, because, at the time, the methods were computationally too intensive.

Each ensemble \hc\ distribution was Gaussian fitted as was demanded by the submission format of the challenge.
However, it is important to note that the distributions are far from Gaussian as discussed in \cite{Denzel20b}.

For each rung, model ensembles were generated for all 16 single lenses and for groups of multiple lenses (four sets of four lenses) using GLASS and Pixelens.
These submissions have the suffixes \texttt{Single} and \texttt{Multi} respectively.

In Rung~1, all GLASS models use $p_r=8$ but single lenses have $sp=5$, and multi-lenses use $sp=1$.
In Rung~2, GLASS single lens models have a higher resolution using $r_p=10$ and $sp=5$, while multi-lenses use $r_p=8$ and $sp=1$.
For Rung~3, the resolution of GLASS models was increased as high as was computationally feasible to $r_p=12$ for the submission \texttt{glassSingleHiRes}. 
The submission \texttt{glassSingleLowRes} used the standard $r_p=8$. 
Both submissions further resolved the central pixel with $sp=3$.

Additionally, in Rung~1 \texttt{glassCherrypick} is a multi-lens analysis using a subset of four lenses for which the individually modelled arrival-time surfaces and mass maps subjectively appeared to be unproblematic (e.g., no additional images and a clean arrival time surface).
In Rung~2, \texttt{glassSynthFiltered} used the aforementioned source mapping algorithm to select models from the \texttt{glassMulti} ensemble which best reproduced the lensed images.

\subsection{Rathnakumar team} \label{glafic_team}
\begin{center}
\small
S. R. Kumar, H. Chand
\end{center}
The main motivation of the team was to understand to what accuracy and precision $H_0$ can be constrained through simple analytical modelling, constrained by point image positions and flux ratios. To this end, the team modelled the TDLMC Rung 0, Rung 1 and Rung 2 systems using {\sc Glafic} software \citep{Oguri2010}. In general, the mass distribution of the lensing galaxy was modelled as singular isothermal ellipsoid along with a shear component (SIE + $\gamma$). In Rung~1, some double lens systems were found to overfit ($\chi^2 << 1$). Thus, the team replaced SIE by singular isothermal sphere (SIS) along with a shear component (SIS + $\gamma$). All the Rung~2 systems were modelled as SIE + $\gamma$, except for one system for which this model was found to result in catastrophic failure. The exceptional case was modelled as singular isothermal ellipsoid without any shear component (SIE only).

The astrometry of the lensed quasar images and the center of the lensing galaxy were measured from the provided \hst\ drizzled image for each system using `imexam' task in IRAF. The astrometric coordinates were assigned an uncertainty of $0\farcs{}02$. The fluxes of the lensed quasar images were also measured through aperture photometry using the same IRAF task from \hst\ drizzled image. From these fluxes, the absolute flux ratio was computed for each lensed quasar image with respect to the brightest image. These flux ratios were each assigned a sufficiently large uncertainty of $0.2$ (e.g., for quads, three flux ratio values were considered), in order to accommodate for factors such as intrinsic quasar variability, microlensing induced variability, etc. Parity constraints were inferred for the lensed quasar images based on the arrival time order and the configuration, in case of quadruple lenses. The team used the velocity dispersion and relative time delay values provided along with their uncertainties as constraints during the modelling. The fitting process was done using standard procedure by implemented in {\sc Glafic}. The background cosmology was fixed to $\Omega_m$ = 0.27, $\Omega_\Lambda$ = 0.73, and $w$ = -1. Source and lens redshifts were fixed for each system according to the provided values. The measured $H_0$ for each system was taken to be that which corresponded to the best fitting model. The 1-$\sigma$ uncertainty of $H_0$ was inferred by fixing it at different values around the measured value and marginalizing all the model parameters to minimize $\chi^2$ and noting the range where $\Delta \chi^2 < 1$, with respect to the value for the best fitting model. The error bars in positive and negative directions were averaged. To include the line of sight effects for Rung~1 and Rung~2 systems, 2.5\% was added in quadrature to the $H_0$ uncertainty. The team submitted only the results for those systems where $H_0$ was constrained to better than $20$~\hcuint. The remaining systems were flagged as failure. The team also submitted results filtered according to cutoff values of $15$~\hcuint and $10$~\hcuint to see what effect these selections have on the TDLMC performance metrics. In order to combine all the $H_0$ estimates from the individual systems into one global value for a rung, the team did  a simple weighted average.   

\newpage
\subsection{H0rton team} \label{NN}
\begin{center}
\small
J. W. Park, Y.-Y. Lin
\end{center}
The H0rton team automated the lens modelling using a Bayesian neural network (BNN), a method pioneered by \cite{hezaveh2017fast}. The BNN-inferred lens model posterior was then propagated into $H_0$ inference. Readers are referred to the accompanying method paper
\citep{park2020large} for more details. The implementation of the H0rton pipeline is available in the form of the open-source Python package {\sc H0rton}.\footnote{\url{https://github.com/jiwoncpark/h0rton}} 

Given the drizzled image of each lens system, the BNN predicted the posterior PDF over a power-law elliptical mass model (PEMD) parameters, the source position, and the half-light radius of the \sersic\ lens light (for computing the velocity dispersion likelihood). The posterior PDF was parameterized as a mixture of two Gaussians with full covariance matrices, informed by the results of 
\citet{wagner2020hierarchical} that the parameter recovery improved with this form of the posterior in comparison to the single uncorrelated Gaussian originally adopted by \citet{hezaveh2017fast}.  

The training set for the BNN consisted of 200,000 images. The assumed lens mass and lens light profiles were identical to those used to generate the TDLMC data of Rung~1 \&~2, i.e., PEMD and elliptical \sersic, respectively. The AGN host light, however, was assumed to follow an elliptical \sersic\ profile in order to keep the parameterization simple. The predictive model parameters in the training set were assumed to be independently distributed, aside from selecting the magnification to be greater than 2 in order to ensure significant lensing signal. The approximate range of each parameter was inferred from the Rung~1 dataset and confirmed by visual inspection on the Rung~3 images. For the PSF convolution, the simulation rotated among the 16 drizzled PSF maps provided in Rung~1. The PSF information was fed to the BNN only via the convolved image and the network was expected to process the deconvolution internally. Non-drizzled images or PSF maps were not used. The training set was generated using the team's open-source Python package {\sc Baobab},\footnote{\url{https://github.com/jiwoncpark/baobab}} which wraps around the {\sc Lenstronomy} package \citep{lenstronomy}.

The combined cosmographic likelihood was the product of the likelihoods of the time delays and the line-of-sight velocity dispersion with the nuisance parameters, i.e, the external convergence, kinematic anisotropy, and the BNN-inferred model parameters, marginalized out. The velocity dispersion was modelled assuming a spherical power-law mass profile and a Hernquist lens light to solve the spherical Jeans equation, as done by \cite{suyu2010dissecting}. The kinematic computations were performed with {\sc Lenstronomy}. Samples from the cosmographic likelihood were obtained via MCMC sampling with \textsc{Emcee} \citep{ForemanMackey:2012p12693}. Note that, in contrast to the traditional forward modelling approach, the pixelwise image likelihood was never directly modelled. Instead, the BNN-inferred posterior entered the MCMC integration as a prior over the lens model parameters at the $H_0$ inference stage.

It was discovered during the analysis procedure that, when the BNN-predicted source position and lens model were directly used to solve the lens equation, the predicted number of images often did not agree with the data. These cases were traced to sources very close to the caustic, for which the precision requirements on the source position tended to be very high \citep[see e.g.][]{birrer2019astrometric}. The BNN-inferred posterior was placing significant weight on models that did not produce the correct number of images. To alleviate this discrepancy, the image positions were manually estimated from the images and fed in as additional data into the MCMC sampling pipeline. A Gaussian likelihood of the image positions, when appended to the MCMC sampling objective, iteratively brought the BNN-inferred lens model closer to one that yielded the observed image positions.

The H0rton team joined the challenge late and only made a blind submission to Rung 3. The open-box datasets of Rungs 1 and 2 that were available at the time, however, informed the team's approach.

\vspace{0.5cm}
\section{Analysis of Rung~1 and Rung~2 submissions}\label{sec_analysis}

To summarize the input data used by each ``Good'' team, we present the information in Table~\ref{sum_table}. A summary of the computation and investigator time invested in the challenge is given in Table~\ref{hours}. A brief analysis of the results of the submissions is presented in this section.

\begin{table}
\centering
\caption{Summary table of input data.} \label{sum_table}
\resizebox{8.5cm}{!}{
\begin{tabular}{lccc} 
		\hline
	Team	 & point sources & extended source & kinematics \\ \\
		\hline
	Student-T &  Yes & Yes & No\\
	EPFL  &  Yes & Yes & Yes\\
	Freeform &  Yes & No & No\\
	Rathnakumar &  Yes & No & Yes\\
	H0rton &  Yes & Yes & Yes\\
                 \hline
\end{tabular}
}
\begin{tablenotes}
\small
\item Note: $-$ Table summarizes the input data as used by the ``Good'' team. In addition, all teams use time delays and redshifts, and simulated \hst\ images to constrain the deflector.
\end{tablenotes}  
\end{table}

\begin{table}
\centering
\caption{Summary of computation and investigator time.} \label{hours}
\resizebox{8.5cm}{!}{
\begin{tabular}{ccc} 
		\hline
	Team	 & CPU time (hours)   &   investigators time (hours) \\ \\
		\hline
	Student-T & $15,400$ & $48$ \\
	EPFL  &  $500,000$ & $1,700$ \\
	Freeform &  $5,000$ & $-$ \\
	Rathnakumar &  $-$ & $-$ \\
	H0rton &  $-$ & $-$ \\
                 \hline
\end{tabular}
}
\begin{tablenotes}
\small
\item Note: $-$ Estimated CPU  and investigator time spent for TDLMC by the teams who provided them.
\end{tablenotes}  
\end{table}

\subsection{Basic statistics}\label{subsec_statistics}
\begin{figure*}
\centering
\includegraphics[width=0.8\linewidth]{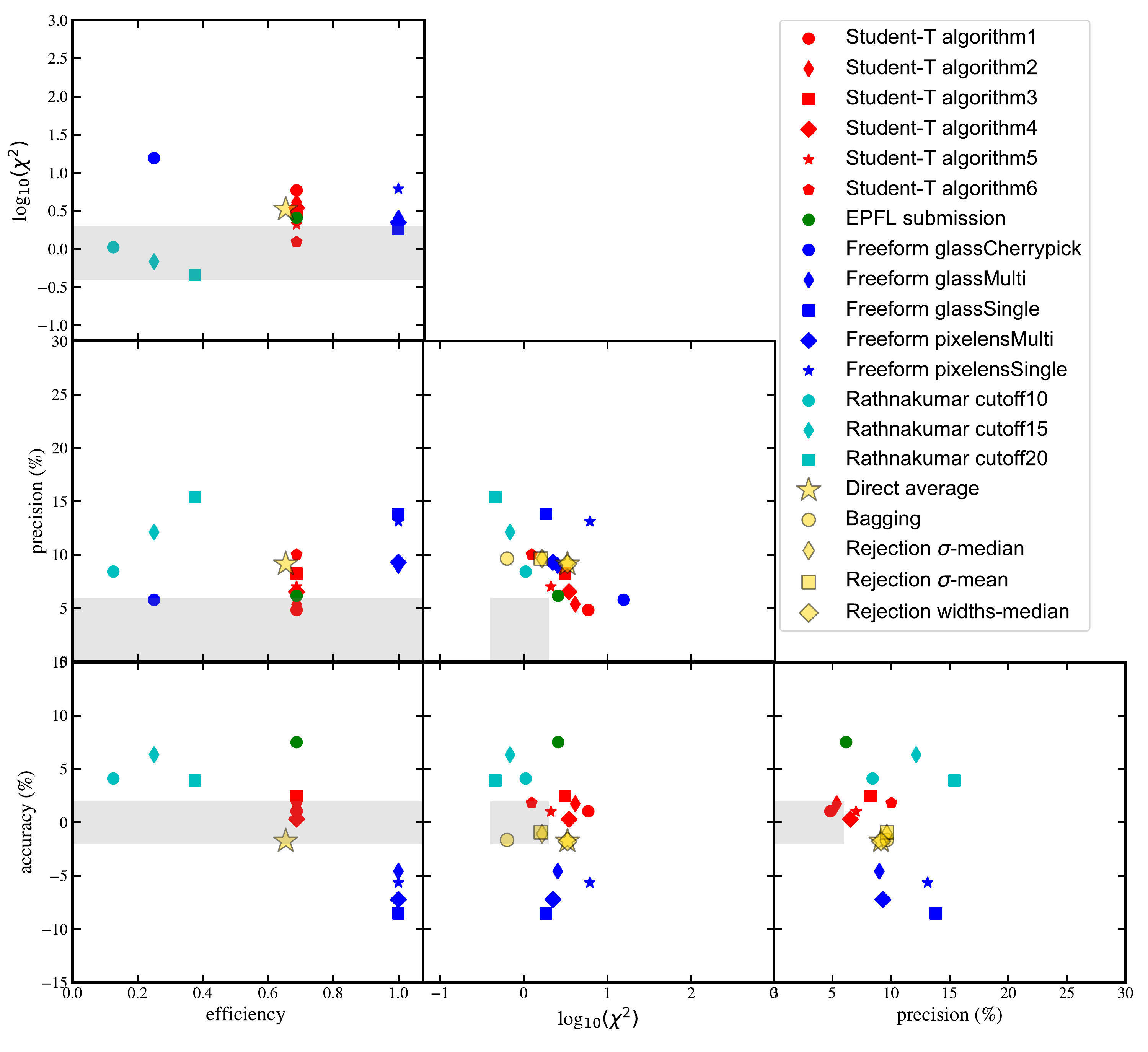}
\caption{
Results for TDLMC Rung~1, showing the 4 metrics for all the submissions using different algorithms, together with the combined metrics shown as yellow points. The $f$, $\chi^2$, $P$ and $A$ are defined in Section~\ref{subsec_expectation}. The gray regions in each plot bracket the expected performance of the metrics. Note that we did not set a target performance for the {\it efficiency} ($f$) metric; the gray regions in the left three panels is drawn only for the other metrics. The last four combined metrics have either reconstructed its sample or rejected the outliers, thus the efficiency metrics are also not considered.
}
\label{fig:rung1_metric}
\end{figure*}

\begin{figure*}
\centering
\includegraphics[width=0.8\linewidth]{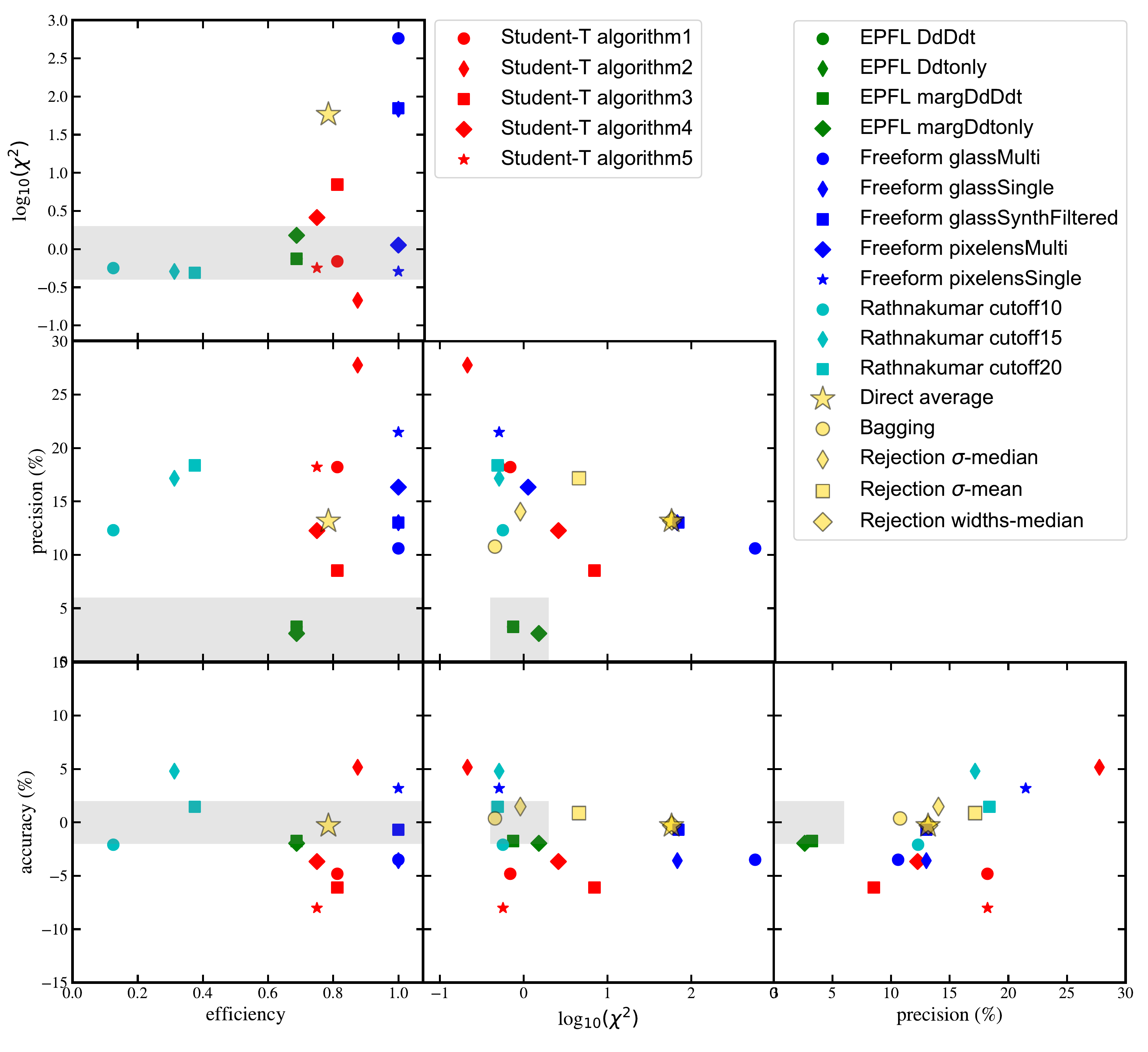}
\caption{
Same as Figure~\ref{fig:rung1_metric}, but for Rung~2's results. To demonstrate the improvement of the Student-T team's result after using the correct file (see Section~\ref{Student-T-correct}), figure also shows post-blind submissions labeled by the hollow markers. We note that the {\it combined metrics}, i.e., yellow points, does not include the results by post-blind algorithms.
}
\label{fig:rung2_metric}
\end{figure*}

In this section, we give an overview of the performance of the blind submissions to Rung~1 and Rung~2. As described in Section~\ref{subsec_expectation}, four metrics are used to perform a synthetic evaluation of the submissions, even though we encourage teams to carry out more detailed studies. The metrics of each submission for Rung 1 and 2 are shown in Table~\ref{table_metrics}. Note that the ``Good'' teams were allowed to adopt multiple methods based on different algorithms and submit multiple results for each rung. The metrics plots by each submission are shown in Figures~\ref{fig:rung1_metric} and \ref{fig:rung2_metric}.

\begin{figure*}
\centering
\begin{tabular}{cc}
\includegraphics[width=0.5\linewidth]{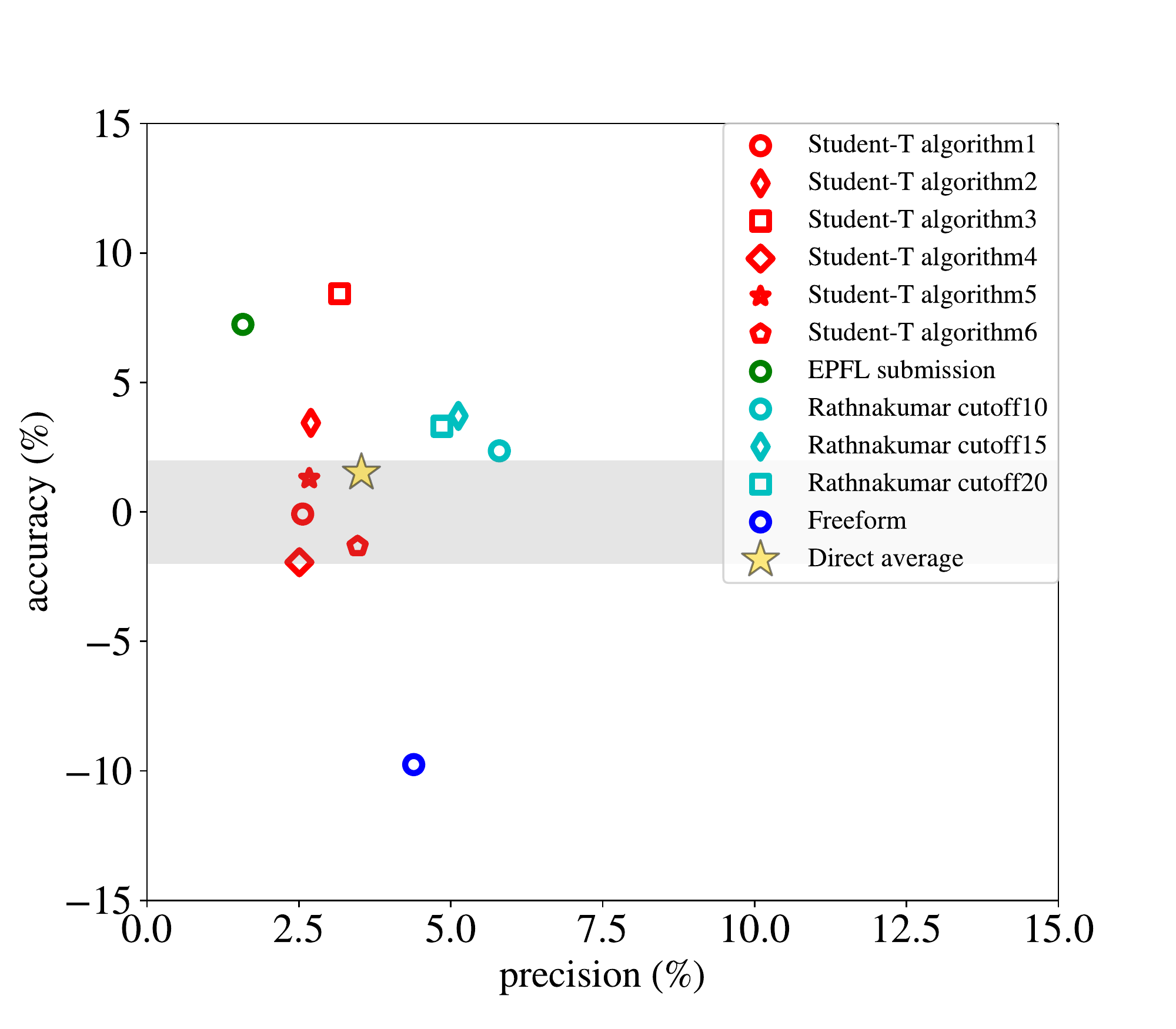}&
\includegraphics[width=0.5\linewidth]{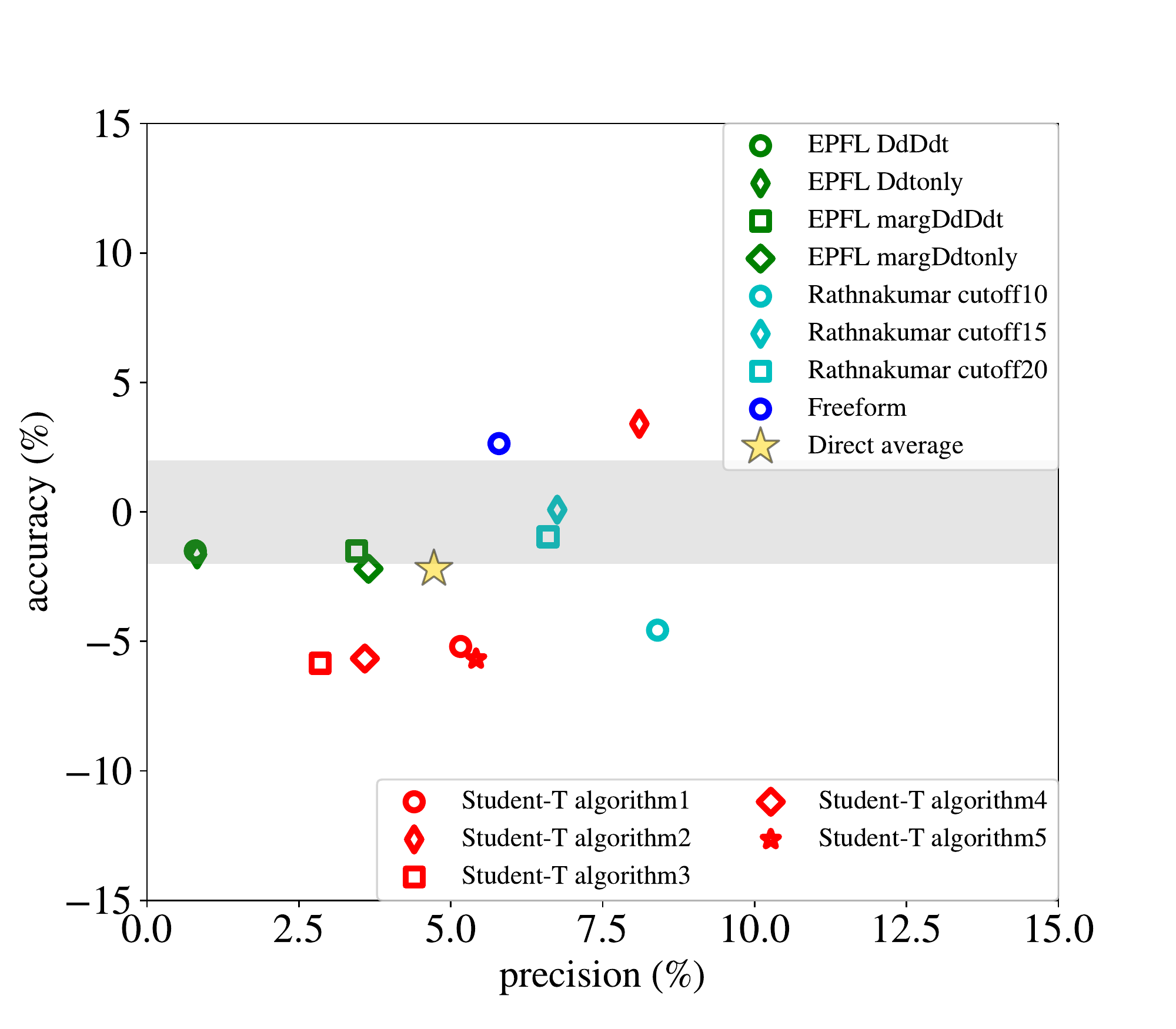}\\
\end{tabular}
\caption{
Results for TDLMC Rung~1 (left) and Rung~2 (right), based on the {\it overall} \hc\ submissions by each algorithm. The Freeform team submitted the {\it overall} \hc\ values after unblinding, although it is based on a straightforward average of blind submissions.
}
\label{fig:rung12_overall_metric}
\end{figure*}

``Good'' teams including Student-T, EPFL and Rathnakumar also estimated and submitted the {\it overall} \hc, which is their best estimation using the combination of the lens systems analyzed in each rung. The Freeform team also submitted the {\it overall} \hc\ values after unblinding, although it is based on a straightforward average of blind inferences. Following Eq.~\eqref{eq:precision} and \eqref{eq:accuracy}, we calculate the metrics of precision and accuracy using the values of these overall  \hc\ and show them in Figure~\ref{fig:rung12_overall_metric}. Note that overall \hc\ is a joint inference from the combination of the multiple lens systems; thus, the precision metric value should be, in principle, decreased by the square root of the volume of the analyzed lensed systems (i.e., $\sqrt{N}$), compared to Figures~\ref{fig:rung1_metric} and \ref{fig:rung2_metric}. The combination of multiple systems could also in principle allow teams to flag and reject outliers, thus reducing the impact of overly complicated systems, i.e., those for which the modelling tool or data quality is insufficient.

Furthermore, we investigated whether there is ``wisdom in the crowd'' by considering metrics combined across \hc\ submissions for Rung~1 and Rung~2. We considered the following strategies:
\begin{itemize}
    \item {\it Direct average}: of all the submission of \hc\, without weighting;
    \item {\it Bagging}: For each lens in one rung, we compute the mean \hc\ across all the submissions and estimate the uncertainty via bootstrap resampling. The result is taken as the \hc\ inference for each lens system. Then, we combine \hc\ inference across all the lens systems in the rung to compute the metrics;
    \item {\it Rejection $\sigma$-median}: We combine the entire \hc\ submissions in one rung to do the bootstrap resampling. We remove the outliers before inferring the averaged metrics using the following criteria. In each bootstrap seed, we calculate the median \hc\ (\hc$_{\rm , median}$) and reject the outliers by $|H_{0 {\rm , median}} - \tilde{H}_{0~i}| /\delta_i>3$;
    \item {\it Rejection $\sigma$-mean}: Similar to  {\it rejection $\sigma$-median}, we remove the outliers in each bootstrap seeding using the  \hc\ weighted mean value \ (\hc$_{\rm , mean}$) by $|H_{0 {\rm , mean}} - \tilde{H}_{0~i}| /\delta_i>3$;
    \item {\it Rejection widths-median}: Similar to previous rejection methods, we use the widths of the \hc\ distribution in each bootstrap drawing (i.e., $W_{H_0}$, which is the half width of $16\%-84\%$ confidence interval in \hc\ distribution) and remove the outliers in the bootstrapped sample by $|H_{0 {\rm , mean}} - \tilde{H}_{0~i}| / W_{H_0}>3$.
\end{itemize}
The combined metrics are shown in Table~\ref{table_metrics} and Figures~\ref{fig:rung1_metric} and~\ref{fig:rung2_metric}. These values can be considered as the combined performance of the entire ``Good'' teams in each rung. As expected, we find that the points of these averaged metrics are in the center of the cloud of the submission by the ``Good'' teams. It is also encouraging that the ensemble averages show no evidence of bias, even though they are a little off the precision target. We note that these combined metrics are inferred after the unblinding in our TDLMC, but they are based on blind submissions. In future blind challenges, this kind of combined metrics could be built in from the start. We note that the {\it averaged metrics} are only introduced to help to ``guide the eye'' to evaluate if there is ``wisdom in the crowd''. This is not a common practice in current research on this topic. Furthermore, the combined metrics are not representative and overweighting certain methods since different teams had different number of submissions.

A few trends emerge from these plots, discarding Student-T submission to Rung~2, and EPFL submission to Rung~1 for reasons discussed in the next subsections. First, most methods seem to have a realistic assessment of their uncertainties, landing on or close to the $\chi^2$ target. Second, the methods constrained only by point source position and fluxes tend to produce significantly larger uncertainties than the target precision. Only the method using the full extent of the surface brightness of the host galaxy and the ancillary data hits the precision target. This trend can be confirmed by Table~\ref{metrics_combined_sample}, in which the combined metrics of {\it precision} and {\it accuracy} are calculated in Rung~2 based on the algorithms using different levels of information. This finding is encouraging even though not surprising: using more data yields more precise results. Also encouraging is that even in the more challenging Rung~2 all the methods - including Student-T post blind - hit the accuracy target. Unexpectedly, the accuracy in Rung~1 seemed to have been less than in Rung~2. 
The improved accuracy in Rung~2 is likely due to the fact that the ``Good'' teams learned from Rung~1's results to improve their algorithms and identify bugs in the codes.

\begin{figure}
\centering
\includegraphics[width=1.0\linewidth]{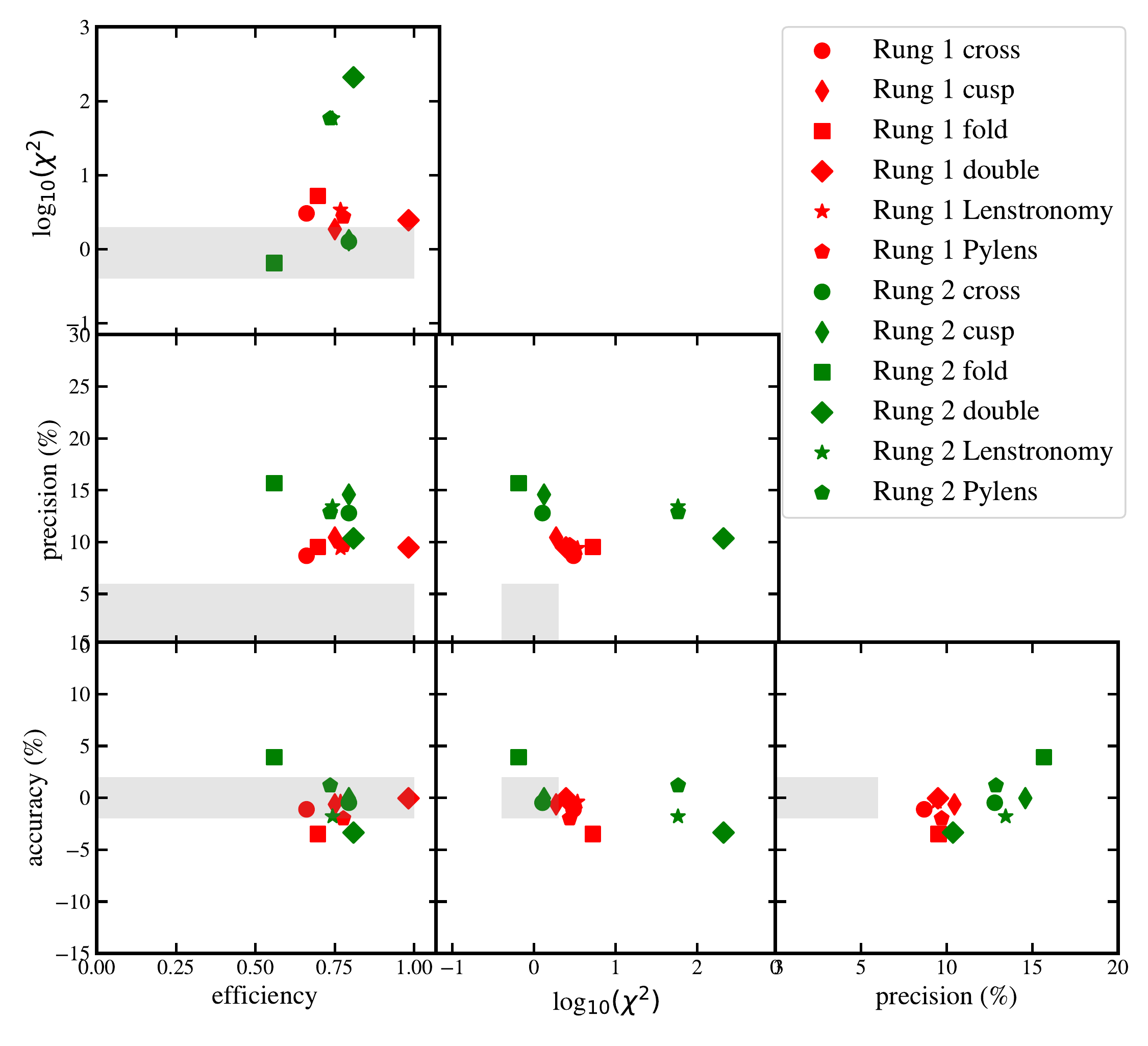}
\caption{
Figure illustrates the metrics of Rung~1 and Rung~2 according to different categories of lens systems using the entire submissions. Note that in Rung 2, the goodness (i.e., $\chi^2$) is overwhelmingly dominated by the four {\it double} systems in the Freeform glassMulti's submission. Because these four double systems are simulated by {\sc Lenstronomy} and {\sc Pylens} evenly, the corresponding $\log_{10}(\chi^2)$ in Rung 2 is significantly larger than the other ones. The larger goodness by Freeform team is an artifact due to the prior that \hc\ is between $50$ and $90$~\hcuint. The values which lie close to $50$ have low error estimates (cut off at 50), which results in very high $\chi^2$ value.
}
\label{fig:catagory_metric}
\end{figure}

To understand if the performance of the lens modelling is different between different lens configurations (i.e., {\it cross}, {\it cusp}, {\it fold} and {\it double}) and simulating codes (i.e., {\sc Lenstronomy} and {\sc Pylens}), we categorize the entire submissions and compare their metrics directly by plotting them together in Figure~\ref{fig:catagory_metric}. Interestingly, there is no significant evidence of difference between the different configurations (e.g., doubles and quads), which is an echo of the recent study by~\citet{Birrer2019_1206} that the precision of the cosmographic measurement with the doubly imaged AGNs could be comparable to those of quadruply imaged ones. Of course, this result should not be overinterpreted as the additional information content of the quads may just be not apparent in the configuration and regimes studied here, but relevant in other situations where for example the mass distribution is more complicated or the data quality is not as good, or the uncertainties are smaller. One potential explanation for the similarity is that the quads considered here are fairly more symmetric than the quads of the TDCOSMO collaborations, likely as a result of the selection function that favors systems with large ellipticity and shear since they have the highest cross-section for quads {\it v.s.} doubles. Symmetric quads have typically shorter time delays and less radial leverage when compared to more asymmetric ones, and thus provide weaker constraints on the Hubble constant. For all these reasons, the similarity between quads and doubles found in this challenge does not imply that they are equally efficient in reality. Also, the metrics are indistinguishable if we consider the {\sc Lenstronomy} and {\sc Pylens} samples separately. This is true even if we restrict the comparison to the submissions by Student-T and EPFL teams, who used {\sc Lenstronomy}. The lack of significant ``home advantage'' is consistent with the fact that the difference of the simulated images between {\sc Lenstronomy} and {\sc Pylens} is below the noise level (see Figure~\ref{fig:sim-images}).

Due to the limitations of Rung~3, as discussed in Section~\ref{sec:rung3}, we present the Rung~3 results in Appendix~\ref{app:rung3}.

\begin{table}
\centering
    \caption{Metrics of blind submission for Rung~1 and Rung~2.}\label{table_metrics}
     \resizebox{8.5cm}{!}{
     \begin{tabular}{llcccc}
     \hline
     Team & algorithm & $f$  &  $\log(\chi^2)$ & $P (\%)$ & $A (\%)$ \\
     &\\
     \hline\hline
     \multicolumn{6}{c}{metrics of Rung~1}\\
     \hline
Student-T & algorithm1 & 0.688 &  0.771 &  4.834 &  1.049 \\
Student-T & algorithm2 & 0.688 &  0.615 &  5.374 &  1.752 \\
Student-T & algorithm3 & 0.688 &  0.493 &  8.237 &  2.492 \\
Student-T & algorithm4 & 0.688 &  0.541 &  6.533 &  0.293 \\
Student-T & algorithm5 & 0.688 &  0.324 &  7.019 &  1.005 \\
Student-T & algorithm6 & 0.688 &  0.094 &  10.036 &  1.825 \\
EPFL  & submission & 0.688 &  0.411 &  6.169 &  7.512 \\
Freeform & glassCherrypick & 0.250 &  1.193 &  5.785 &  -22.847 \\
Freeform & glassMulti & 1.000 &  0.406 &  9.002 &  -4.570 \\
Freeform & glassSingle & 1.000 &  0.264 &  13.812 &  -8.516 \\
Freeform & pixelensMulti & 1.000 &  0.349 &  9.299 &  -7.220 \\
Freeform & pixelensSingle & 1.000 &  0.790 &  13.123 &  -5.632 \\
Rathnakumar & cutoff10 & 0.125 &  0.024 &  8.429 &  4.112 \\
Rathnakumar & cutoff15 & 0.250 &  -0.164 &  12.137 &  6.337 \\
Rathnakumar & cutoff20 & 0.375 &  -0.339 &  15.419 &  3.932 \\     
\hline
 \multicolumn{2}{c}{Rung~1 combined metrics} &   &   &    &   \\ \hline
   {\it Direct average} &   & 0.654 &  0.522 &  9.140 &  -1.745\\
  {\it Bagging}   & &  & -0.199 & 9.646 & -1.644 \\
  {\it Rejection $\sigma$-median}   & &  & 0.219 & 9.639 & -1.081 \\
  {\it Rejection $\sigma$-mean}   & &  & 0.205 & 9.649 & -0.920 \\
  {\it Rejection widths-median}   & &  & 0.522 & 9.147 & -1.779 \\
     \hline\hline
     \multicolumn{6}{c}{metrics of Rung~2}\\
     \hline 
Student-T & algorithm1 & 0.812 &  -0.161 &  18.215 &  -4.811 \\
Student-T & algorithm2 & 0.875 &  -0.672 &  27.764 &  5.161 \\
Student-T & algorithm3 & 0.812 &  0.845 &  8.531 &  -6.096 \\
Student-T & algorithm4 & 0.750 &  0.414 &  12.267 &  -3.663 \\
Student-T & algorithm5 & 0.750 &  -0.247 &  18.225 &  -8.014 \\
EPFL  & DdDdt & 0.688 &  -0.127 &  3.260 &  -1.740 \\
EPFL  & Ddtonly & 0.688 &  0.180 &  2.635 &  -1.957 \\
EPFL  & margDdDdt & 0.688 &  -0.127 &  3.260 &  -1.740 \\
EPFL  & margDdtonly & 0.688 &  0.180 &  2.635 &  -1.957 \\
Freeform & glassMulti & 1.000 &  2.762 &  10.603 &  -3.496 \\
Freeform & glassSingle & 1.000 &  1.834 &  13.010 &  -3.580 \\
Freeform & glassSynthFiltered & 1.000 &  1.847 &  13.017 &  -0.683 \\
Freeform & pixelensMulti & 1.000 &  0.053 &  16.335 &  17.095 \\
Freeform & pixelensSingle & 1.000 &  -0.293 &  21.480 &  3.187 \\
Rathnakumar & cutoff10 & 0.125 &  -0.249 &  12.304 &  -2.090 \\
Rathnakumar & cutoff15 & 0.312 &  -0.293 &  17.166 &  4.797 \\
Rathnakumar & cutoff20 & 0.375 &  -0.311 &  18.382 &  1.461 \\
\hline
 \multicolumn{2}{c}{Rung~2 combined  metrics}  &   &   &    &   \\ \hline
  {\it Direct average} &   & 0.785  & 1.765 &  13.154 &  -0.309  \\
  {\it Bagging}  & &  &  -0.343 &  10.768 & 0.372 \\
  {\it Rejection $\sigma$-median}   & &  &  -0.040 &  14.041 & 1.481 \\
  {\it Rejection $\sigma$-mean}  & &  &  0.660 &  17.170 & 0.870 \\
  {\it Rejection widths-median}   & &  &  1.769 &  13.187 & -0.279 \\
    
\hline
 \multicolumn{6}{c}{Rung~2 post-blind submissions, see Sec~\ref{Student-T-correct}}\\
\hline 
Student-T & algorithm1 & 0.938 &  -0.421 &  15.492 &  -5.969 \\
Student-T & algorithm2 & 1.000 &  -0.873 &  26.844 &  6.396 \\
Student-T & algorithm3 & 1.000 &  0.317 &  6.591 &  0.056 \\
Student-T & algorithm4 & 1.000 &  -0.162 &  11.805 &  4.330 \\ 
     \hline\hline
      \end{tabular}}
\begin{tablenotes}
\small      
\item Note: $-$ Table summaries the metrics of the blind submission for Rung~1 and Rung~2, together with the post-blind submissions by Student-T team (see Section~\ref{Student-T-correct}).
\end{tablenotes}  
\end{table}      

\begin{table}
\centering
\caption{Summary of the {\it precision} and {\it accuracy} by combining algorithms based on different level of information used to constrain the models in Rung~2.} \label{metrics_combined_sample}
 \resizebox{8.5cm}{!}{
\begin{tabular}{lcr} 
		\hline
	Combined fitting algorithm & Precision $(\%)$ & Accuracy $(\%)$ \\ \\
		\hline
	Everything &  $2.9$ & $-1.8\pm0.4$ \\ \\
	\hline
	Extended Source:\\
	~~~{\it blind submissions only}  & $11.4$ &  $-2.7\pm1.0 $ \\
	~~~{\it blind + post-blind}  & $12.8$ &  $-1.2\pm0.8 $ \\
	~~~{\it only post-blind for Student-T}  & $10.1$ &  $0.02\pm0.69 $ \\ \\
	\hline
	Point Sources & $15.2$ & $2.5\pm1.4 $ \\
                 \hline                 
\end{tabular}}
\begin{tablenotes}
\small
\item Note: $-$ ``Everything'' calculates the metrics combining the algorithms that adopted point sources, extended source, and kinematics. ``Extended Source'' combines the results of the algorithms that utilized the lensed arc information in the lens modelling. ``Point Sources'' combines the ones which use only point sources but not lensed arcs. For cases with post-blind submissions explained in the text we report all the permutations of blind and post-blind combinations.
\end{tablenotes}  
\end{table}

\subsection{Lessons form Rung~1 and Rung~2}\label{subsec_lesson0}

The first important lesson is that the independent teams have come up with several independent techniques, including novel ones. As described above, the underlying assumptions of the techniques vary greatly, and so does the amount of information used by each technique and the flexibility of the models. As often the case in astrophysics, finding the right balance between too little and too much flexibility in the models is difficult yet vital to obtain accuracy and precision. Too little flexibility may lead to bias or underestimated error bars. Too much flexibility may lead to unphysical solutions or unnecessary inflation of the error bars. The level of flexibility directly ties to another major obstacle to precision, lensing degeneracies. One way in which degeneracies can be quantified is by pulling multiple solutions from different families of models, and analyzing the variance within that ensemble~\citep[see e.g.,][]{Gomer19, Saha2000}.

The second important lesson is that most methods seem to produce reasonable estimates of their uncertainties. In Rung~1 virtually all methods produced acceptable $\chi^2$ metric distributions, while in Rung~2 the submissions that returned an answer for every system (i.e., high efficiency) sometimes paid the price in the sense that they underestimated their uncertainties.

The third important lesson is that more information translates to higher precision. Therefore, if one wishes to extract high precision from time delay measurements, it is crucial to use all the information available, not just the positions of the point sources (or their flux). However, an important caveat is that information content by itself does not necessarily guarantee accuracy if the modelling technique is not sufficiently flexible, as discussed above. Rung~1 and Rung~2 provide a useful test, but much remains to be done to explore the right degree of flexibility.

After unblinding Rung~1, the EPFL team discovered that small systematic uncertainties in the position of the multiply imaged quasars at the level of a fraction of a pixel could introduce a noticeable bias in the inference given the precision of the time delays. Thus, in Rung~2, the EPFL team introduced nuisance parameters to describe this uncertainty and marginalized over it. The effect is evident by comparing their blind results in Rung~1 and Rung~2. This is an example of the importance of modelling technique flexibility to ensure accuracy, and the fourth key lesson from Rung~1 and Rung~2 is that astrometric precision needs to be commensurate with the time delay precision. As discussed by \citet{birrer2019astrometric} the requirements can be at the level of milli-arcseconds if the time delay is known to percent precision. For \hst-like images, the requirements correspond to a small fraction of a pixel, a challenging requirement for point sources superimposed on an extended and unknown source. It is thus important to consider explicitly this source of uncertainty and marginalize it, transforming a potential source of bias into a decrease in precision.

\subsection{Notes about Student-T's submissions for Rungs~2 and~3}\label{Student-T-correct}
\begin{figure*}
\centering
\includegraphics[width=0.9\linewidth]{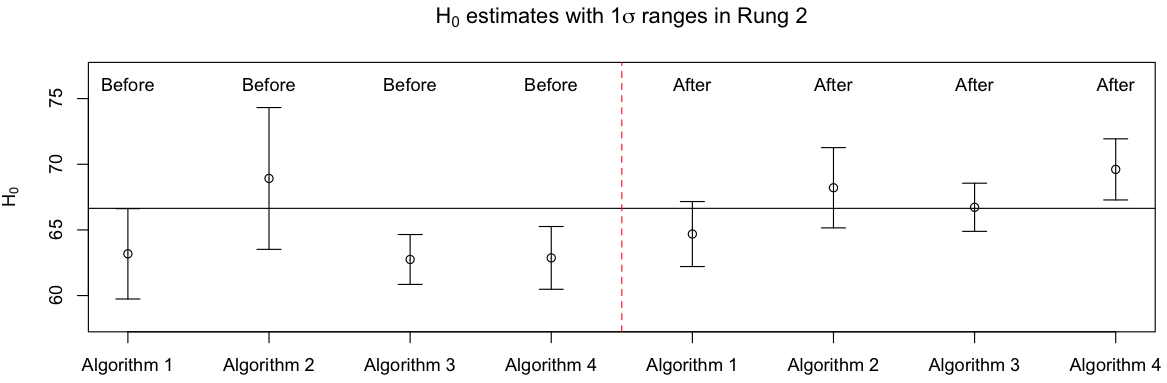}
\caption{
Post-blind improvement of the Student-T team's results using the correct PSF file for Rung~2 submissions, without changing any code or algorithm. This correction removes any bias in the inference and improves slightly the precision.
}
\label{fig:Student-T_imporve}
\end{figure*}

After unblinding, it was discovered that in Rung~2 (and Rung~3) the Student-T team used the non-drizzled PSF, drizzled lens image, and drizzled noise map, owing to clerical errors. The team's unblinded (post) analyses show that this mismatch was the main source of biases in the blinded analysis. In Figure~\ref{fig:Student-T_imporve}, we find that the Rung~2's result after using the correct file is much improved. The corresponding metrics of the post analysis are also given in Table~\ref{table_metrics}. We stress that these post-submissions only corrects the input file; the modelling algorithms remain unchanged. These post-submissions are not used while calculating the combined (i.e., averaged) metrics.

\section{Limitations of Rung~3, including post-unblinding discoveries}\label{sec:rung3}

Rung~3 was inconclusive because of the limitations of the procedure used to construct the lenses for this rung. We discuss here some of the limitations of the hydrodynamical simulations used to construct Rung~3. The ``Evil'' team was aware of some of them while constructing the challenge, while others only became apparent post-unblinding. We introduce them in the following subsection.

\subsection{Limitations known before unblinding}

The main known limitations of the simulations pre-unblinding are twofold. 

\begin{figure}
\centering
\includegraphics[width=1.0\linewidth]{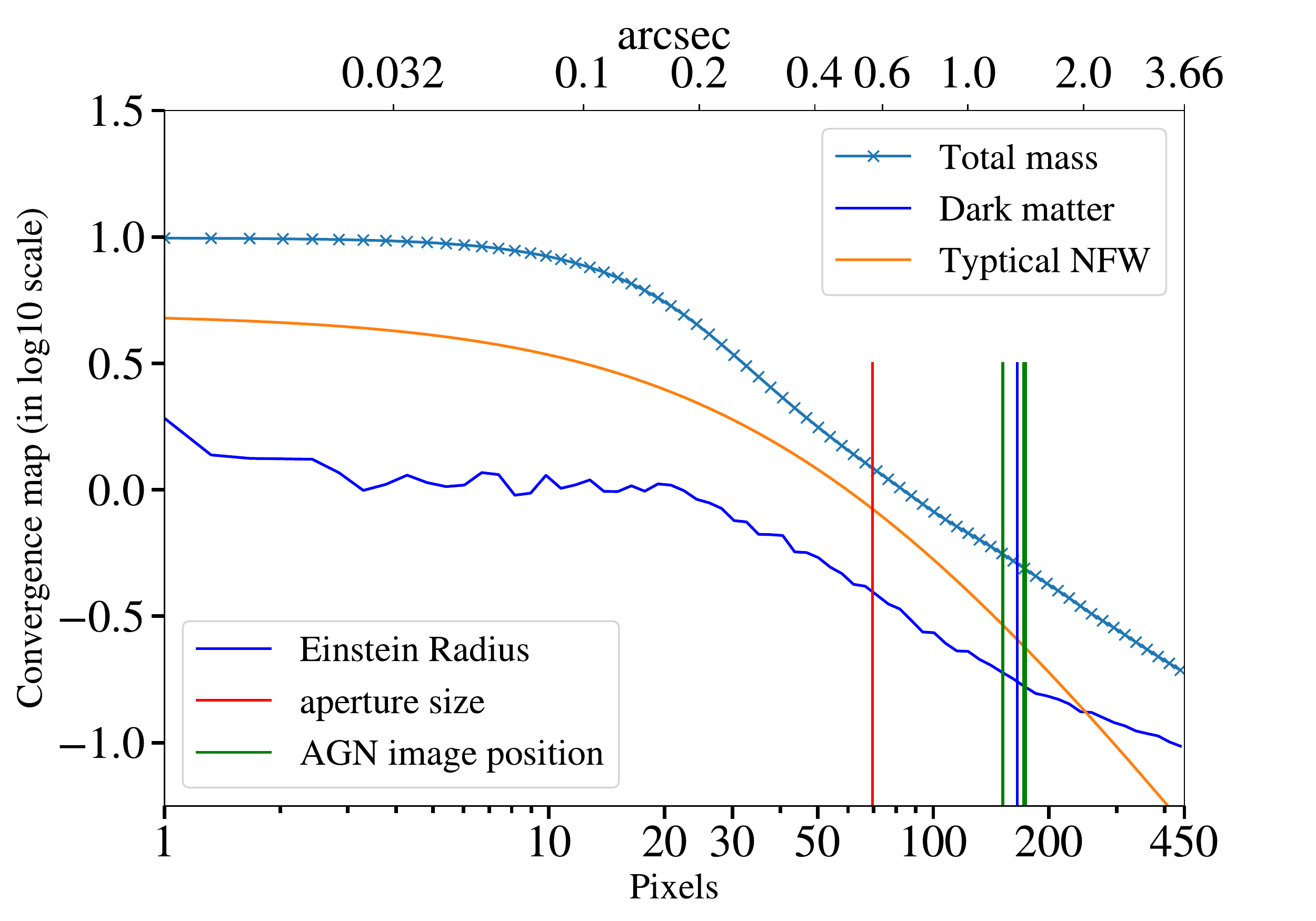}
\caption{
Mass profile of a typical deflector in Rung~3, illustrating the unphysical core in the central regions and the departure of the dark matter halo from a standard \citep{Navarro1997} form.
}
\label{fig:profile}
\end{figure}

First, the resolution of the simulations we used is insufficient to describe the inner regions of early-type galaxies. This is illustrated in Figure~\ref{fig:profile}, where we show a typical mass profile, decomposed in dark and total mass.
The total mass profile has a core of approximately $0\farcs1$, about half a kpc at the redshift of our sources. We also note that the adopted numerical simulations have softening lengths of 200$-$700 pc, which have partially contributed to the core sizes in these simulated galaxies.
Despite that some cored massive elliptical galaxies have been found~\citep{Thomas2016} and could be produced in highly accurate dynamical simulations~\citep{Rantala2018}, they are unlikely to be present in real lens galaxies with mass like Rung 3 ones.  A recent detailed analysis of the mass density profiles of massive lens galaxies \citep{Shajib2020} in terms of stars and dark matter halos shows that the dark matter halo is well described by a ``cuspy'' unperturbed \citet{Navarro1997} halo and that the population of the lens galaxies' total mass density profile is close to a power-law profile (within $\sim$5 per cent near the Einstein radius). Whereas simulations have made a lot of progress in reproducing massive elliptical galaxies, \citet{Shajib2020} show that they still fall short in simultaneously reproducing the mass density profile and the dark matter fraction of real galaxies at the level of detail needed for this test.

The main evidence against cores is from the search for central images of gravitational lenses themselves. The central slope of the mass density profile controls the magnification of the central image. The fact that the central image is almost always absent in galaxy scale lenses (not in clusters-scale lenses), is a strong argument against cores.
For example, radio observations ~\citep[e.g.,][]{Rusin2001,Keeton2003, Winn2004,Boyce2006,Zhang2007,Quinn2016} usually present a non-detection of the `central' lensed image, which gives an upper limit of the core (<5$\sim$100 pc). Likewise, in the TDCOSMO project, which models the high-resolution lensed AGN images based on HST observations, the fifth image has not been detected, although as we show below at optical/infrared wavelengths contamination by the deflector light limits the sensitivity.

\begin{figure*}
\centering
\begin{tabular}[b]{c}
\small (a) Mock lensed arcs for Rung~2 \\
\includegraphics[trim = 55mm 40mm 40mm 30mm, clip,width=1.\linewidth]{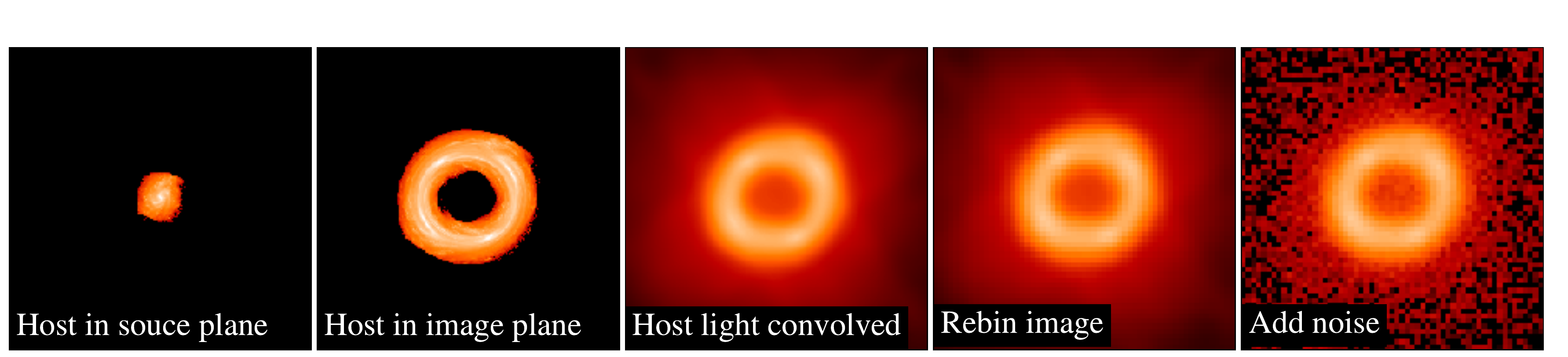}\\ \\
\end{tabular}

\begin{tabular}[b]{c}
\small (b) Mock lensed arcs for Rung~3 \\
\includegraphics[trim =  55mm 40mm 40mm 30mm, clip,width=1\linewidth]{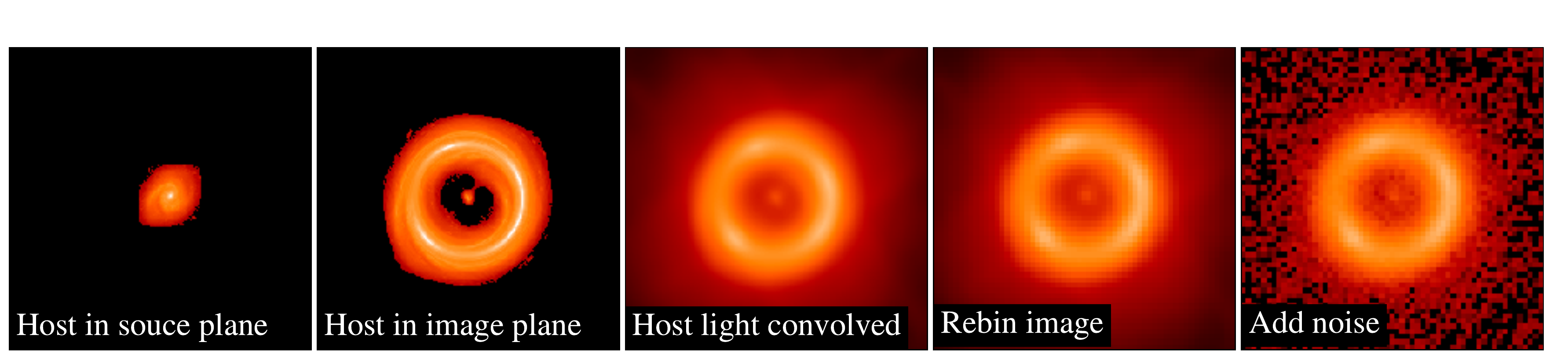}\\
\end{tabular} 
\caption{
Mock images of the lensed arcs in the simulations of Rung~2 and Rung~3. Due to the unphysical core of Rung~3 deflector's mass profile, the lensed arcs appear the feature of `central' lensed image (bottom-second plane).  However, after entire simulation process this feature is not distinct anymore and would be overwhelmed by deflector light. 
}
\label{fig:lens_arcs}
\end{figure*}

A simple ``gedanken experiment'' shows that the cores present in the Rung~3 simulations are unphysical, and therefore justifies our caution interpreting them. As shown in Figure~\ref{fig:lens_arcs}, Rung~3 predicts a central image while Rung2 does not. Unfortunately, in the optical and near infrared such central image is difficult to disentangle from the light of the deflector.

In contrast, if we could perform the observations of the Rung~3 systems in the radio, assuming the multiply-imaged point source is radio loud, the test would be conclusive.
The mean value of the magnification of the central source $\mu_{\rm c}$ for Rung~3's simulations is $\sim0.032$, which is significantly larger than the upper limit level reported by~\citet[][i.e., $\mu_{\rm c}<0.001$]{Keeton2003}. 
Furthermore, we calculated the ratio between the $\mu_{\rm c}$ and the magnification of the standard lensed point sources ($\mu_{\rm bright}$) and found that the mean value of $\mu_{\rm bright}/\mu_{\rm c} $ is $\sim192$, which is inconsistent with the values reported in the literature \citep[][i.e., >2500, >1000, >10000, respectively]{Boyce2006, Zhang2007, Quinn2016}. These results indicate that the core feature in Rung~3's simulations is not realistic.

The second argument to use Rung~3 with caution is that since simulations do not match perfectly the mass profile of real massive elliptical galaxies, as shown by Figure~\ref{fig:profile}, generalizing the results of such a test is always going to be complicated. 
For example, if the modelers were to assume the mass density profile to be cuspy in the inner regions and thus do not match the cores in the simulations, would this be a problem in analyzing real galaxies, which should be cuspy? The recent study by \citet{Enzi2019} shows that without kinematic information, departures from a single power-law (in this case, in the form of a core) can lead to a bias on the inference of \hc\ of up to 25\%. A similar concern about the realism of simulations is illustrated by \citet{Xu2017}, who analyzed Illustris simulations and showed that the simulations do not match exactly the detailed properties of real galaxies in terms of central dark matter fraction and slope of the mass density profile (see also~\citet{Wang2020, Shajib2020}.

These limitations were known to the ``Evil'' team while designing the challenge. The ``Evil'' team considered these limitations a ``necessary evil'', to be kept in mind in the interpretation of the results.
Indeed, when simulating the mock images, the ``Evil'' team was aware that the Rung~3's lensed arcs demonstrated the fifth image feature, compared to Rung2's simulation, see Figure~\ref{fig:lens_arcs}. However, this feature is not detectable in the simulated images due the contamination from the deflector light (the fifth image flux ratio is $<0.1\%$, compared to the deflector light).

In the end, the benefit of knowing the three-dimensional ``truth'' for a complex system was considered to outweigh the downside of the system not being fully realistic.

Future challenges may want to pursue some form of empirically-driven models (perhaps based on observations of local massive elliptical galaxies) until the fidelity of simulations improves significantly.

\subsection{Limitations discovered post-unblinding}

Additional limitations were discovered post-unblinding thanks to collaborative efforts by the ``Evil'' and ``Good'' teams. However, these limitations do not necessarily invalidate the mock data or introduce a major bias to ``Good'' team's inference of \hc.

\subsubsection{Substructure and dynamics}

In Rung~3, 12/16 simulations dynamically bound substructures (i.e., satellite halos) were identified and removed before producing the lensing quantities.
This procedure renders the kinematics inconsistent with the lensing quantities, because the motion of the stars and gas was precomputed based on the full mass distribution including substructure. Substructure accounts for approximately 1\% of the total mass at the relevant scales, so this is not a large effect, but can potentially introduce a bias at the percent level when combining lensing and kinematic tracers.

\subsubsection{Halo truncation}
For computational reasons, only the particles within the virial radius ($R_{200}$) or twice the virial radius were considered when projecting the mass distribution to calculate lensing quantities. This introduces two main outcomes. First, not taking into account mass beyond $R_{200}$ may introduce a negative mass-sheet transform, biasing $H_0$ below the percent level. Second, the spherical truncation at $R_{200}$ does not follow the isodensity contours of the mass profile, introducing an artificial shear~\citep{Van_de_Vyvere2020}. At this radius, the artificial shear created by the truncation is small and may bias $H_0$ by less than 1 percent. Both truncation effects (i.e., artificial shear introduction and negative mass-sheet bias) have low amplitude for truncation at the virial radius. They then may introduce a small bias on the $H_0$ inference but should not be the major cause of bias in Rung~3 results.

\vspace{0.5cm}

\section{Discussion and implications for future work}\label{sec_implications}

First of all, a positive outcome of the challenge is that several teams were able to analyze a sample of 48 lenses, the sample size needed to reach sub-percent precision \citep{Shajib2017}. Analyzing this large sample within the time constraints of the challenge required good teams to apply fast methods as opposed to the more time and resource consuming approaches of state-of-the-art analysis of real data. These fast methods are necessary to make progress, and it is essential to test them as we did in the challenge. We note that even with the fast methods participation to the challenge was labor intensive, and the ``Evil'' team extended the original deadlines set in TDLMC1 by a few months in order to allow more ``Good'' teams to participate.

Rung~1 and Rung~2 demonstrate that current fast lens modelling technology is able to obtain precise and accurate estimates of \hc\ starting from a best guess of the point spread function, when using the information content of \hst-like images. The expected complexity of the lensed host galaxy of the quasar is not an obstacle to the inference, provided that sufficiently flexible models are used to describe the source.  The common practice of reconstructing the PSF starting from an empirical or theoretical best guess and the use of flexible source description is validated by the two rungs and should become the standard in future work.

Astrometry of the point sources from \hst-like images can be a source of bias at the few percent level for extremely precise time delays. Mitigation strategies include adding nuisance parameters to describe the astrometric noise arising from poor sampling, or using higher resolution images, e.g., from adaptive optics or radio interferometers.

The conclusions about modelling the gravitational potential of the deflector are not so clear cut. Encouragingly, the teams performed well when the deflector was described by a simply parametrized analytic forms as in Rung~1 and Rung~2, with no evidence of inaccuracy. As discussed above, and as expected, the fast methods using more information performed better in terms of precision than the ones which used only AGN positions and flux ratios. Rung~3 was helpful in unveiling subtle effects that need to be considered if one wishes to use simulations to test gravitational lens modelling techniques for cosmological inference to high precision. Unfortunately, the same limitations  -- and the known limitations in resolution and realism at the beginning of the challenge -- make it difficult to draw conclusions based on it. More work is needed on this front, and it will require either much higher resolution simulations than the ones adopted here or more advanced computational techniques to calculate the lensing quantities. Alternatively, a future challenge could find a way to generate high precision and realistic models, perhaps inspired by empirical data on local massive elliptical galaxies.

\vspace{0.5cm}
\section{Summary and conclusions} \label{sec_summary}

In this paper, we described the main results of the time delay lens modelling challenge. We first revealed some of the details of the construction of the simulated datasets that were kept blind during the challenge. Second, we gave a brief description of the methods followed by the ``Good'' teams to do the inference. Third, we described a number of limitations of Rung~3, including some numerical effects discovered post-unblinding that preclude inferences at the percent level required for this challenge. These limitations make Rung~3 difficult to interpret but are reported here with the aim to inform future challenges. Finally, we presented an overview of the performance of the methods against 4 metrics (precision, accuracy, efficiency, goodness of fit). 

The main conclusions, based on Rungs 1 and 2, can be summarized as follows:
\begin{itemize}
    \item Each team came with fundamentally different methods to study a large sample of systems. In particular, methods constrained only by point-like images and using either analytic or free-form models, a novel Bayesian technique assuming a locally Gaussian Fermat potential, and modelling similar to current cosmographic analyses. A Bayesian Neural Network approach has also been applied on unblinded data. Several teams developed fast methods that allowed them to analyze 48 lenses within the duration of this challenge ($\sim1-2$ years). This is a much larger number of systems per investigator time than the current state-of-the-art models, that so-far requires of order $\sim1$ year per system (not considering the process of collecting ancillary data and analyzing the lens environment).
    \item The fast methods applied to this challenge estimate their uncertainty appropriately, yielding error bars that are statistically comparable with the departure from the truth. 
    \item The fast methods that exploit the full information content of the data achieve higher precision than the ones that only utilize lensed quasars positions and fluxes to constrain the models.
    \item The fast methods based on full image reconstruction can meet the target precision ($6\%$ per system) and accuracy ($2\%$) when analyzing mock images based on complex sources and starting with a guess of the point spread function.  
    \item Astrometric requirements on the position of the point sources can be stringent and difficult to meet for high precision time delay measurements, given the Hubble Space Telescope point spread function and pixel size. Biases arising from the poor sampling of the PSF can be avoided by modelling the astrometric noise explicitly.
\end{itemize}

As far as Rung~3 is concerned, one generic problem was known before the challenge, i.e., if simulations do not reproduce real galaxies at the percent level precision in gravitational potential, it is difficult to generalize the outcome of the challenge. A good example of this issue is the finite resolution of cosmological hydrodynamical resolution, which introduces features like cores that are unlikely to be present in real systems. A spherical redistribution of cusp to core would not itself affect lensing observables, but it would change kinematic and other properties. If modelers assume that galaxies are cuspy, and do not detect the core in the simulations, what does it mean for real galaxies? The following additional and more subtle effects were identified post-unblinding. 

\begin{itemize}
    \item The kinematics of the particles in the simulations must be consistent to sub-percent level with the gravitational potential generated by the lensing data products given to the ``Good'' teams. Removing substructures or other parts of the simulation when generating the lensing data may cause internal tension in the data so that the lensing and dynamical probes cannot be combined without bias.
    \item  The standard practice of truncating simulated halos at the virial radius may lead to inconsistencies between the actual Fermat potential and the one computed from truncated maps. Lensing quantities such as the Fermat potential are non local, and the kernel mapping convergence into potential is logarithmic. Therefore, in order to avoid biases in Fermat potential at the few percent level, one has to include all particles well beyond the virial radius and carefully consider the shape of the truncation.
\end{itemize}

In recent years, a number of works have investigated the systematic uncertainties in time-delay cosmography~\citep[e.g.,][]{Schneider2013, Birrer2016, Sonnenfeld2018, Kochanek2020,Millon2019}. However, it is difficult to make a quantitative comparison between our results and those in the literature because the uncertainties depend strongly on the assumptions and methods used.

To conclude, this work shows that blind challenges on simulated data are a powerful tool to study and characterize a method, alongside blind and independent analysis of real datasets \citep{Millon2019}. The results obtained from this first time delay lens modelling challenge are encouraging, in the sense that accurate and precise \hc\ can be derived blindly even in the presence of complex sources and unknown PSF. However, our results also demonstrate that much work remains to be done before we can have conclusive end-to-end tests based on simulations. First, state-of-the-art modelling methods exploiting the full information content of the data need to speed up so that even larger simulated datasets can be analyzed within a practical time frame to explore a variety of more complicated configurations. For example, the EPFL team that used all the information employed 500,000 CPU hours and 1,700 hours of investigator time, almost a full year equivalent. This is significantly less time than currently employed per lens by H0LiCOW or STRIDES. However, the challenge was single plane and by design simpler in terms of satellites and perturbers along the line of sight than real lenses. So, in order to analyze samples of order 100-1000 lenses with increased complexity, further speed-ups are necessary. 

Second, improvements in numerical simulations of massive elliptical galaxies and the calculation of their lensing properties are needed before they can be used to perform lens modelling challenges to percent level precision.

\section*{Acknowledgments}
We thank Kenneth C. Wong, Sherry H. Suyu for useful suggestions and supports.

T.T. acknowledges support by the Packard Foundation in the form of a Packard Research Fellowship.
T.T. and C.D.F. acknowledge support by NSF through grant ``Collaborative Research: Toward a 1\% Measurement of The Hubble Constant with Gravitational Time Delays'' AST-1906976.
This project has received funding from the European Research Council (ERC) under the European Union's Horizon 2020 research and innovation programme (grant agreement No 787886). 
C.D.F. and G.C.-F.C. acknowledge support for this work from the National Science Foundation under Grant Numbers AST-1312329 and AST-1907396.
This work has received funding from the European Research Council (ERC) under the European Union's Horizon 2020 research and innovation programme (COSMICLENS: grant agreement No 787886) and the Swiss National Science Foundation (SNSF).  
A.J.S. was supported by the National Aeronautics and Space Administration (NASA) through the Space Telescope Science Institute (STScI) grant HST-GO-15320. This research was supported by the U.S. Department of Energy (DOE) Office of Science Distinguished Scientist Fellow Program. 
S.V. has received funding from the European Research Council (ERC) under the European Union's Horizon 2020 research and innovation programme (grant agreement No 758853).

Hyungsuk Tak acknowledges Simon Birrer for his sincere and tireless support on implementing \textsc{Lenstronomy}, which has enabled the team's participation in the challenge. Hyungsuk Tak also appreciates Xuheng Ding for his thoughtful comments on the team's Python code submitted to the Evil team, which has significantly improved the methodology (including the discovery of the clerical error). In addition, Hyungsuk Tak acknowledges computational supports from the Institute for Computational and Data Sciences at Pennsylvania State University.  

S. Rathna Kumar and Hum Chand acknowledge financial support from SERB, DST, Govt. of India through grant PDF/2016/003848 during the course of this project. 
S. Hilbert acknowledges support by the DFG cluster of excellence `Origin and Structure of the Universe'.
SV thanks the Max Planck Society for support through a Max Planck Lise Meitner Group, and acknowledges funding from the European Research Council (ERC) under the European Union's Horizon 2020 research and innovation programme (LEDA: grant agreement No 758853).

At the moment of this writing the revised version of the manuscript following the referee's comments, the mock data of TDLMC have already been used as a validating set for subsequent analyses~\citep{Birrer2020}.

\section*{Data Availability}
The data underlying this article are available in the TDLMC website, at \url{https://tdlmc.github.io/}



\bibliographystyle{mnras}


\appendix
\newpage

\section{Details of Rung 3}
\label{app:rung3}

{
\subsection{Illustris simulations}
The first group of simulated galaxies is selected from the Illustris simulation~\citep{Vogelsberger2013, Vogelsberger2014a} with six galaxies at $z=0.4$ and six galaxies at $z=0.6$. All have total dark matter halo masses between $1-2~(10^{13} M_\odot$), and velocity dispersion ranging from $250$~km/s to $320$~km/s. In this challenge, we do not intend to test biases in the most severe cases where the true profiles significantly deviate away from the power-law models. For this reason, our selection was based on the fact that the selected galaxies shall distribute fairly closely around the best-fit general mass-velocity dispersion relation. As a result, the majority of the selected galaxies are not classified as the extreme cases of deviations from power-law mass distributions; the most severe case would result in an underestimate of Hubble constant by $15\%$ \citep[see][]{Xu2016}.

The convergence and potential maps (as well as potential's first and second derivatives) were calculated using netted-mesh based methods through FFT with an isolated boundary condition. All matter distribution of the selected galaxy halo is truncated at R200 with a spherical aperture~\citep{Xu2009}. The results have been cross-checked with the public software GLAMER, which is a ray-tracing code for the simulation of gravitational lenses~\citet{Metcalf2014, Petkova2014}.
In addition, we also calculated the same maps using a mesh-based FFT algorithm, adopting Smoothed-particle hydrodynamics (SPH) kernel to smooth the simulated particles to the mesh. The two sets of results showed expected consistency within the numerical uncertainties.

The velocity maps were calculated on desired meshes; here no smoothing was used. The pixel values of mean velocity and velocity dispersions were weighted by rest-frame SDSS-r band luminosities of stellar particles projected to the pixel.     

\subsection{Zoom simulations}
The second set of simulations is a sample of `zoom' cosmological simulations, which have been previously used in~\citet{Frigo2019}. A `zoom' simulation is a higher resolution re-run of a small part of the cosmological box of a large-scale simulation (like Illustris), called the `parent' simulation. In the set we employed, the parent simulation is a 100 Mpc wide cosmological box simulated with dark matter only~\citep{Oser2010}, and each zoom simulation covers the volume of a dark matter halo (at $z=0$). The simulations were run with a modified version of GADGET2~\citep{Springel2005}, called SPHGAL~\citep{2014MNRAS.443.1173H}, which avoids some of the shortcomings of SPH codes. Unlike the parent, the zoom simulations also include gas, stars and black hole particles. They include models for star formation (based on gas density and temperature), metal enrichment, gas cooling, stellar winds, supernova feedback (Type Ia and Type II), and AGN feedback (using the \citet{2012ApJ...754..125C} model). The spatial resolution (softening length) of the simulation is 200 pc, while the mass resolution (initial mass of gas particles) is $7\times10^5 \, M_\odot$. This is a higher resolution than Illustris, but not high enough to avoid the issues presented in Section~\ref{sec:rung3}. The simulations run from $z=43$ to $z=0$. The sample of simulated galaxies varies in mass, size, dynamical and stellar-population properties. For the TDLMC project, we used snapshots at different redshifts ($0.3 < z < 0.5$) of the four most massive AGN galaxies, which have arcsec-size Einstein radii. More details on the simulation code and on this sample can be found in~\citet{Frigo2019}.

The maps of convergence, lensing potential and its derivatives were calculated with the post-processing ray tracing code {\sc Hilbert}~\citep{Hilbert2007, Hilbert2009}. The whole high resolution region of each simulation, roughly reaching out to twice the virial radius of the galaxy, was fed into the code and used to calculate the lensing maps. The 3D orientation of the galaxy was chosen randomly before the analysis. The kinematic maps were calculated on the same grid as the lensing maps, weighting the line-of-sight velocity of each particle with its R band luminosity.

\subsection{Rung 3 results}
For completeness, we report here the full results of Rung~3. We caution the reader that the interpretation of these results is difficult, because of the limitations and numerical issues described in Section~\ref{sec:rung3}.

The metric of each submission for Rung~3 are listed in Table~\ref{table_metrics_Rung3} and plotted in Figures~\ref{fig:rung3_metric} and~\ref{fig:rung3_metric_2}.

\begin{table}
\centering
    \caption{Metrics of blind submission for Rung 3.}\label{table_metrics_Rung3}
     \resizebox{8.5cm}{!}{
     \begin{tabular}{llcccc}
     \hline
     Team & algorithm & $f$  &  $\log(\chi^2)$ & $P (\%)$ & $A (\%)$ \\
     &\\
     \hline\hline     
     \multicolumn{6}{c}{metrics of Rung~3}\\
     \hline
Student-T & algorithm1 & 0.750 &  0.117 &  15.616 &  -3.803 \\
Student-T & algorithm2 & 0.812 &  -0.583 &  26.226 &  6.221 \\
Student-T & algorithm3 & 0.938 &  0.459 &  8.472 &  1.677 \\
Student-T & algorithm4 & 1.000 &  0.213 &  12.869 &  2.512 \\  
Student-T & algorithm5 & 0.875 &  0.402 &  11.998 &  -11.998 \\
Student-T & algorithm6 & 0.938 &  -0.932 &  26.515 &  3.986 \\
Student-T & algorithm7 & 1.000 &  0.718 &  4.885 &  -5.415 \\
Student-T & algorithm8 & 1.000 &  0.027 &  12.587 &  -2.786 \\
Student-T & algorithm9 & 0.875 &  0.532 &  8.247 &  -7.373 \\
Student-T & algorithm10 & 0.938 &  -0.848 &  15.369 &  4.401 \\
Student-T & algorithm11 & 1.000 &  1.132 &  3.923 &  -5.065 \\
Student-T & algorithm12 & 1.000 &  0.115 &  9.728 &  -1.195 \\
EPFL  & Combined & 0.438 &  0.893 &  4.276 &  -9.963 \\
EPFL  & CombinedDdtOnly & 0.438 &  0.879 &  4.584 &  -9.944 \\
EPFL  & Composite & 0.500 &  1.515 &  2.612 &  -11.302 \\
EPFL  & CompositeDdtOnly & 0.500 &  1.500 &  2.559 &  -11.403 \\
EPFL  & Powerlaw & 0.812 &  0.938 &  2.941 &  -7.016 \\
EPFL  & PowerlawDdtonly & 0.812 &  0.955 &  3.001 &  -6.973 \\
Freeform & glassMulti & 1.000 &  2.464 &  5.106 &  -16.041 \\
Freeform & glassSingleHiRes & 1.000 &  1.954 &  5.809 &  -17.267 \\
Freeform & glassSingleLowRes & 1.000 &  1.401 &  9.632 &  -11.441 \\
Freeform & pixelensMulti & 1.000 &  -0.695 &  18.866 &  7.626 \\
Freeform & pixelensSingle & 1.000 &  -0.226 &  21.637 &  0.542 \\
H0rton & Bayesian neural network & 0.312 &  0.637 &  9.056 &  3.356 \\    
     \hline\hline 
\end{tabular}}
\begin{tablenotes}
\small      
\item Note: $-$ Table summaries the metrics of the blind submission for Rung~3.
\end{tablenotes}  
\end{table}

\begin{figure*}
\centering
\includegraphics[width=0.8\linewidth]{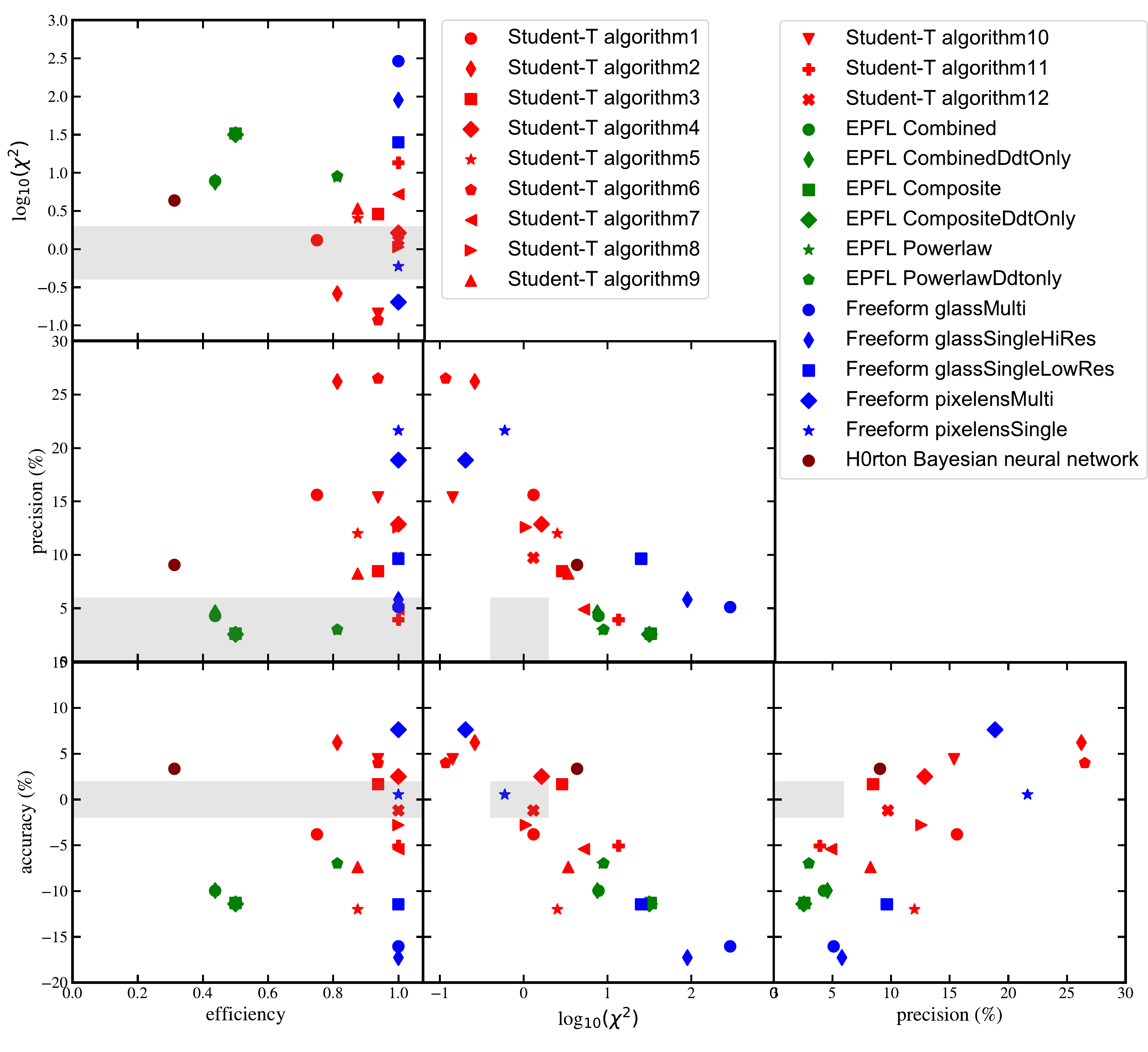}
\caption{
Metrics of Rung~3 blind submissions. Note that Rung~3 was affected by issues described in Section~\ref{sec:rung3} and thus great caution should be taken in interpreting these results.
}
\label{fig:rung3_metric}
\end{figure*}

\begin{figure*}
\centering
\begin{tabular}{cc}
\includegraphics[width=0.5\linewidth]{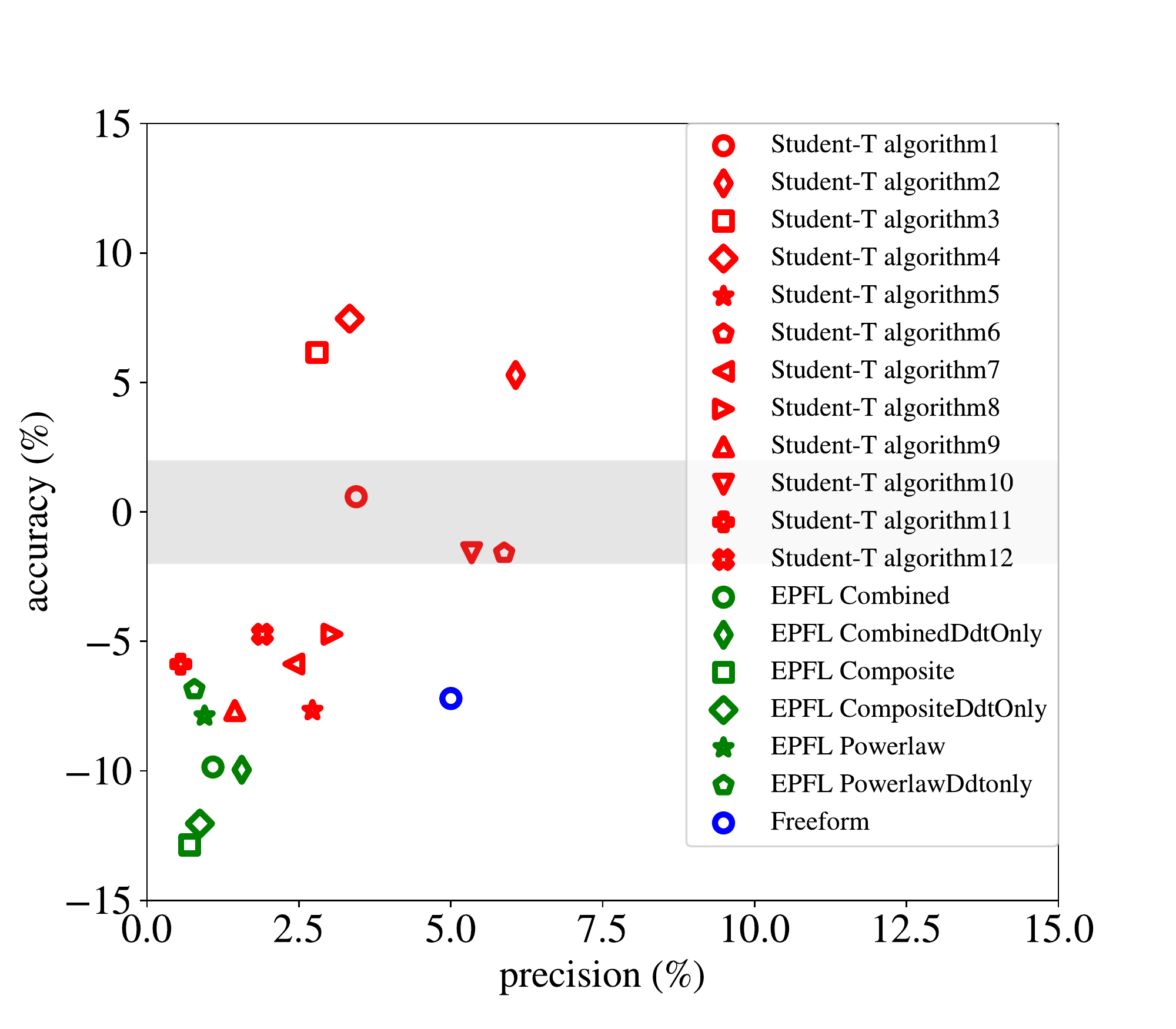}&
\includegraphics[width=0.5\linewidth]{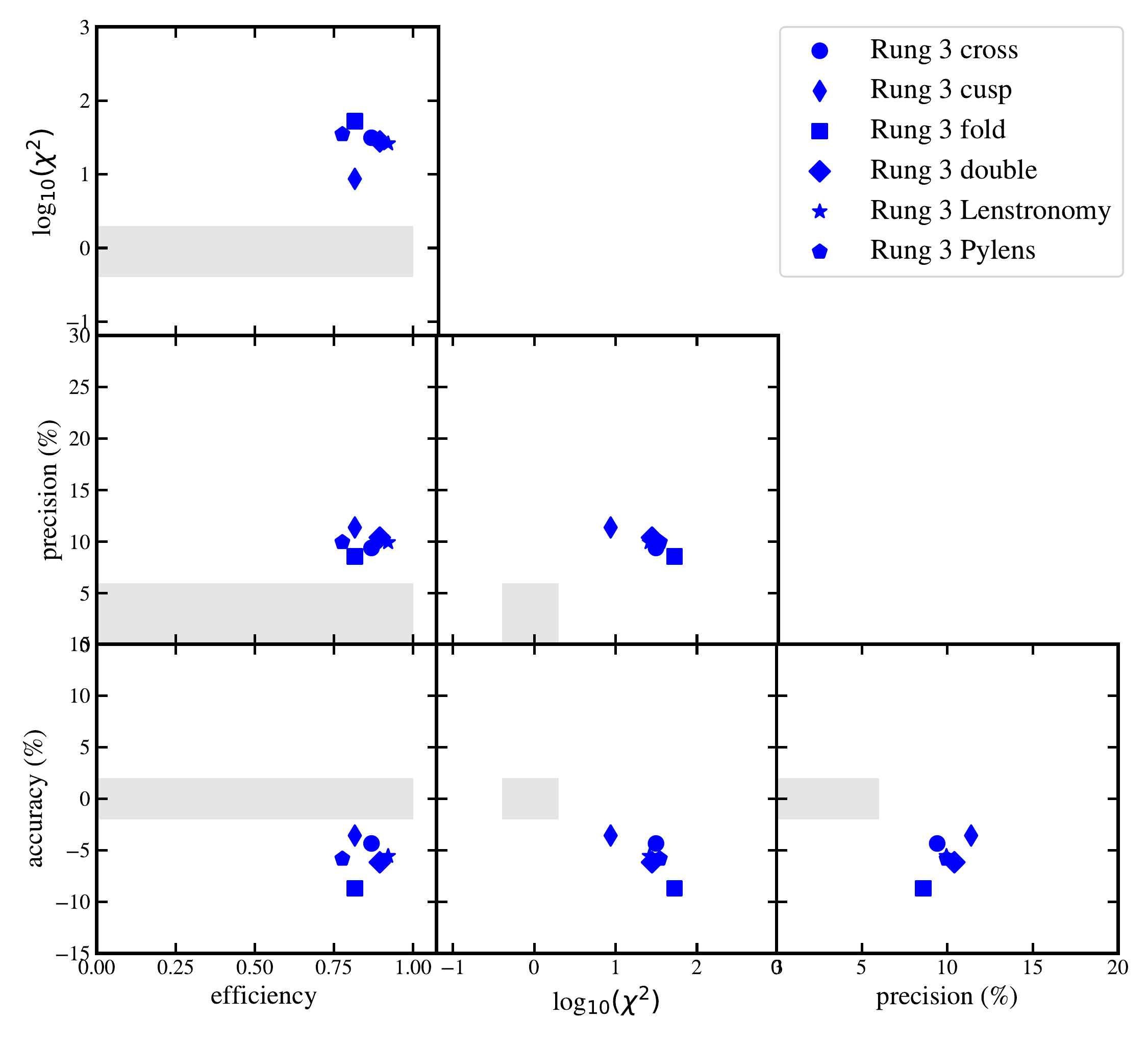}\\
\end{tabular}
\caption{
Panel (left) and (right) is the same as Figure~\ref{fig:rung12_overall_metric} and~\ref{fig:catagory_metric}, separately, but for Rung~3.
}
\label{fig:rung3_metric_2}
\end{figure*}


\bsp	
\label{lastpage}
\end{document}